\documentclass[aps,prl,twocolumn,superscriptaddress]{revtex4-1}
\usepackage[colorlinks=true,bookmarks=false,citecolor=black,urlcolor=black,linkcolor=black,breaklinks]{hyperref} 
\usepackage{breakurl}
\usepackage{sidecap}
\sidecaptionvpos{figure}{t}
\usepackage{amsmath}
\usepackage{graphicx}
\usepackage{esint}
\usepackage{epstopdf}
\graphicspath{{pict/}{./}}
\usepackage{bm}%
\usepackage{breakurl}
\usepackage{sidecap}
\sidecaptionvpos{figure}{t}
\usepackage{amsmath}
\usepackage{graphicx}
\usepackage{esint}
\usepackage{epstopdf}
\graphicspath{{pict/}{./}}
\usepackage{bm}%
\usepackage{titlesec}
\usepackage{amssymb}

\newcounter{Fig}

\begin{document}

\title{Multipolar Conversion Induced Subwavelength High-Q Supermodes with Unidirectional Radiations}
\author{Weijin Chen}
\affiliation{School of Optical and Electronic Information, Huazhong University of Science and Technology, Wuhan, Hubei 430074, P. R. China}
\author{Yuntian Chen}
\email{yuntian@hust.edu.cn}
\affiliation{School of Optical and Electronic Information, Huazhong University of Science and Technology, Wuhan, Hubei 430074, P. R. China}
\author{Wei Liu}
\email{wei.liu.pku@gmail.com}
\affiliation{College for Advanced Interdisciplinary Studies, National University of Defense
Technology, Changsha, Hunan 410073, P. R. China}

\begin{abstract}
The two-mode coupling model with energy splitting and formation of supermodes with different life times has been pervasive in almost every discipline of physics.  We revisit this fundamental model  from a different perspective of multipolar expansions, and manage to reveal a hidden dimension of it, by establishing a subtle connection between two seemingly unrelated properties of Q-factors and far-field angular radiation patterns. We discover that, in both regimes of negative and positive couplings, significant Q-factor enhancement can be attributed to dramatic redistributions of radiations that originate from multipolar conversions from lower to higher orders. Relying on this connection and generalized Kerker effects of interferences among different multipoles, we synchronize both outstanding features of high-Q factor and unidirectional radiation into one subwavelength supermode. The implications of our study are not confined to optics and photonics, and can potentially  shed new light on coupling between resonances of mechanical, phononic, electronic or other hybrid natures.
\end{abstract}
\maketitle

The elegant model of two-mode coupling has been serving as one of the most fundamental frameworks for  different branches of physics and many other interdisciplinary  fields~\cite{SHANKAR_2011__Principlesa,KITTEL_2004__Introduction,DEMTRODER_2018__Atoms,BLATT_2010__Theoretical,HAUS_1983__Waves,Yariv2006_book_photonics,vahala_optical_2004,VUKOBRATOVICH_2018__Fundamentalsa}.
This model is generically related to lots of exotic phenomena including Fano resonances~\cite{Miroshnichenko2010_RMP,LIMONOV_2017_Nat.Photonics_Fano}, electromagnetically induced transparencies~\cite{FLEISCHHAUER_2005_Rev.Mod.Phys._Electromagnetically,HAMMERER_2010_Rev.Mod.Phys._Quantum}, bound states in the continuum~\cite{HSU_Nat.Rev.Mater._bound_2016,FRIEDRICH_Phys.Rev.A_interfering_1985-1,HEISS_Phys.Rev.E_repulsion_2000-1}, scarred states in open systems~\cite{WIERSIG_Phys.Rev.Lett._formation_2006-1,SONG_2010_Phys.Rev.Lett._Improving}, non-Hermitian and topological effects~\cite{MOISEYEV_2011__NonHermitian,FENG_2017_Nat.Photonics_NonHermitian,Hasan2010_RMP,Lu2014_topological}, and so on. For the simplest case of two coupled resonances supported by open and passive resonators, generally two supermodes with different life times would emerge~\cite{HAUS_1983__Waves,Yariv2006_book_photonics} (refer to Ref.~\cite{Supplemental_Material_2} for more details). The longer-lived higher-Q supermode that originates from out-out-phase superposition can locate on the lower- or higher-energy branch, depending on the sign (positive or negative, respectively) of the coupling strength~\cite{FRIEDRICH_Phys.Rev.A_interfering_1985-1,HEISS_Phys.Rev.E_repulsion_2000-1,AWAI_2007_Electron.Commun.Jpn.PartIIElectron._Coupling}. On one hand, the formation of the higher-Q supermode can be intuitively attributed to the radiation loss suppression as a result of far-field destructive interference between the two original states~\cite{HAUS_1983__Waves,Yariv2006_book_photonics,FRIEDRICH_Phys.Rev.A_interfering_1985-1,HEISS_Phys.Rev.E_repulsion_2000-1,Supplemental_Material_2}. On the other hand, however, the description \textit{destructive interference} is itself, to some extent still hazy, but rather we do not know exactly what really happens in the far field that has induced the Q-factor enhancement.

To more accurately grasp the far-field properties of the higher-Q supermode and to provide new insights,  we revisit this fundamental two-mode coupling model in photonic resonators and manage to reveal a hidden dimension of it. Our study here is conducted from a different perspective of multipolar expansions, which serve as a fundamental tool for far-field scattering or radiation analysis~\cite{jackson1962classical,Bohren1983_book,DOICU_light_2006,POWELL_Phys.Rev.Applied_interference_2017,LIU_2018_Opt.Express_Generalized}. In sharp contrast to previous studies that mostly treat  Q-factors and angular radiation distributions as unrelated properties,  here we try to establish a connection between them~\cite{Supplemental_Material_2}. It is discovered that in the anti-crossing regions of strong mode coupling  with both negative and positive coupling strengths, the Q-factor enhancement is intrinsically connected with multipolar conversions from lower to higher orders (see Ref.~\cite{Supplemental_Material_2} for the coupling type classification and extra evidence that such a connection is also manifest in a weak coupling scenario). Based on this subtle connection and generalized Kerker effects of interferences among multipoles of different natures (electric or magnetic) and orders to produce directional radiations~\cite{LIU_2018_Opt.Express_Generalized,Kerker1983_JOSA,jahani_alldielectric_2016,KUZNETSOV_Science_optically_2016}, we manage to synchronise three features of subwavelength mode volume, high-Q factor and unidirectional radiation pattern into one supermode. We show that such subwavelength Kerker supermode can be realized with simple dielectric nanoscale resonators with broken symmetries, which is vitally important for nanoscale lasers, sensors, and single-photon sources.

\begin{figure}[tp]

\centerline{\includegraphics[width=8.8cm]{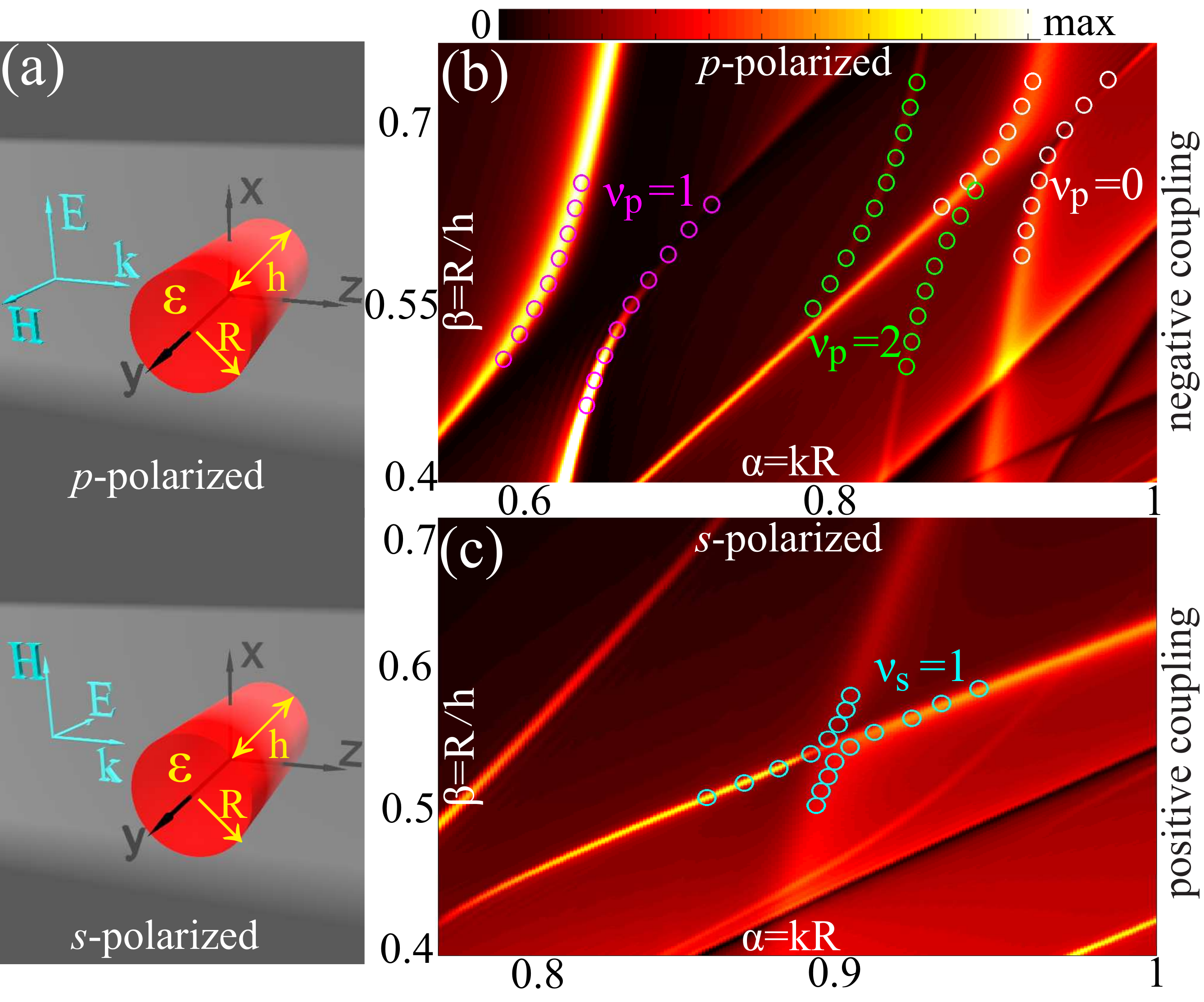}} \caption {\small (a) Schematic of a dielectric rod resonator of height $h$, radius $R$, and relative permittivity $\varepsilon$.
The exciting normally incident plane wave propagating along $\textbf{z}$ can be $\textbf{\textit{p}}$- or $\textbf{\textit{s}}$-polarized. The scattering cross section spectra for the rod  are shown in (b) and (c), for $\textbf{\textit{p}}$ and $\textbf{\textit{s}}$ polarizations respectively. Four strong mode coupling induced anti-crossing regions characterized by different azimuthal quantum numbers are marked, and within each region two supermode branches are pinpointed by circles. For $\textbf{\textit{p}}$ ($\textbf{\textit{s}}$) polarizations, the supermodes locate on the right (left) branches experience significant Q-factor enhancement, indicating negative (positive) coupling strengths.}\label{fig1}
\end{figure} 

\begin{figure*}[tp]
\centerline{\includegraphics[width=16cm]{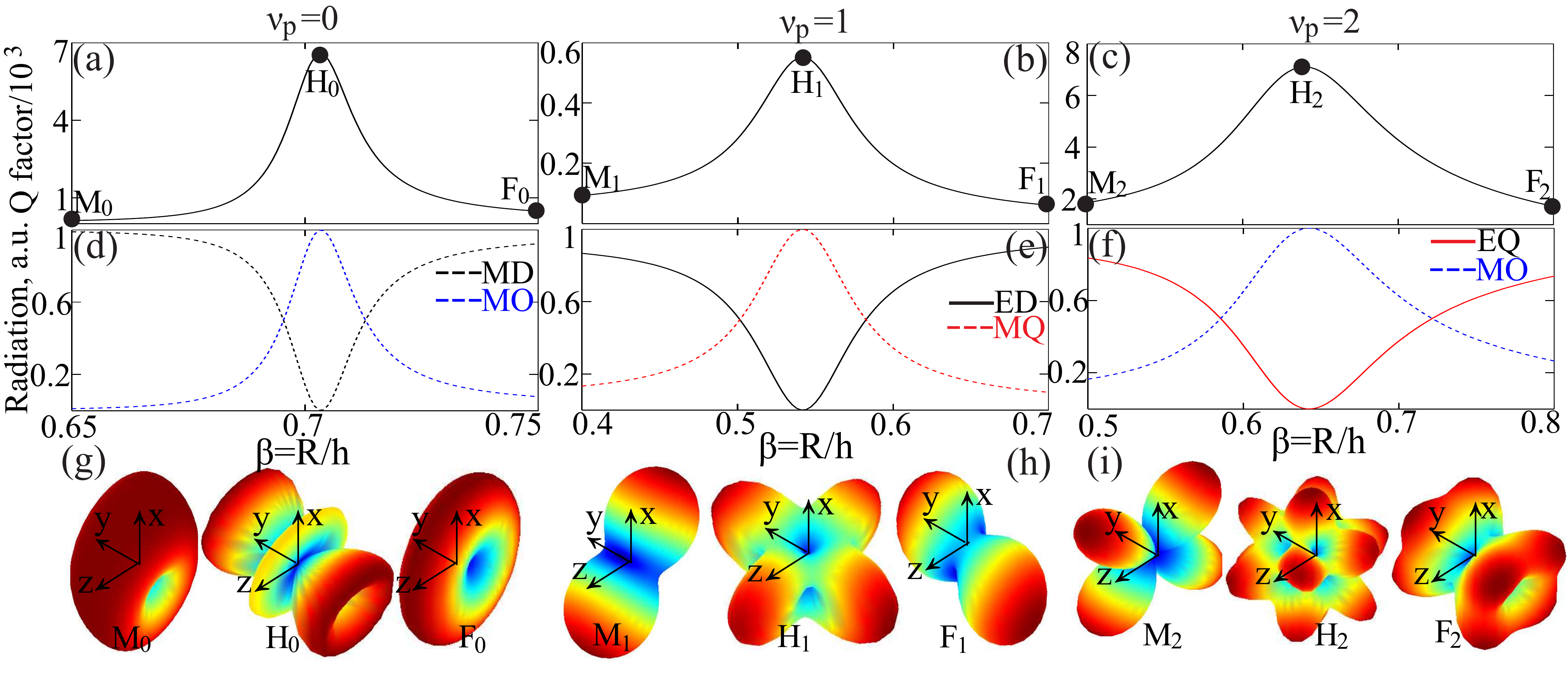}} \caption{\small (a)-(c) Evoluations of Q factors, and (d)-(f) the radiated power from all multipoles that are not negligible, for the supermodes located on the right branches in the three anti-crossing regions marked in Fig.~\ref{fig1}(b), with $\nu_\textbf{p}=0,1,2$ respectively. For each case in (a)-(c), three supermodes are indicated by points $\textbf{M}_{0,1,2}$, $\textbf{H}_{0,1,2}$, and $\textbf{F}_{0,1,2}$, and the corresponding far-field radiation patterns of those supermodes are shown respectively in (g)-(i). Specific parameters for the indicated points are: $\beta_\textbf{M}=0.65,~0.4,~0.5$, $\beta_\textbf{F}=0.703,~0.541,~0.641$, and $\beta_\textbf{F}=0.75,~0.7,~0.8$, for $\nu_\textbf{p}=0,1,2$ respectively.}
\label{fig2}
\end{figure*}

\begin{figure}[tp]
\centerline{\includegraphics[width=7.5cm]{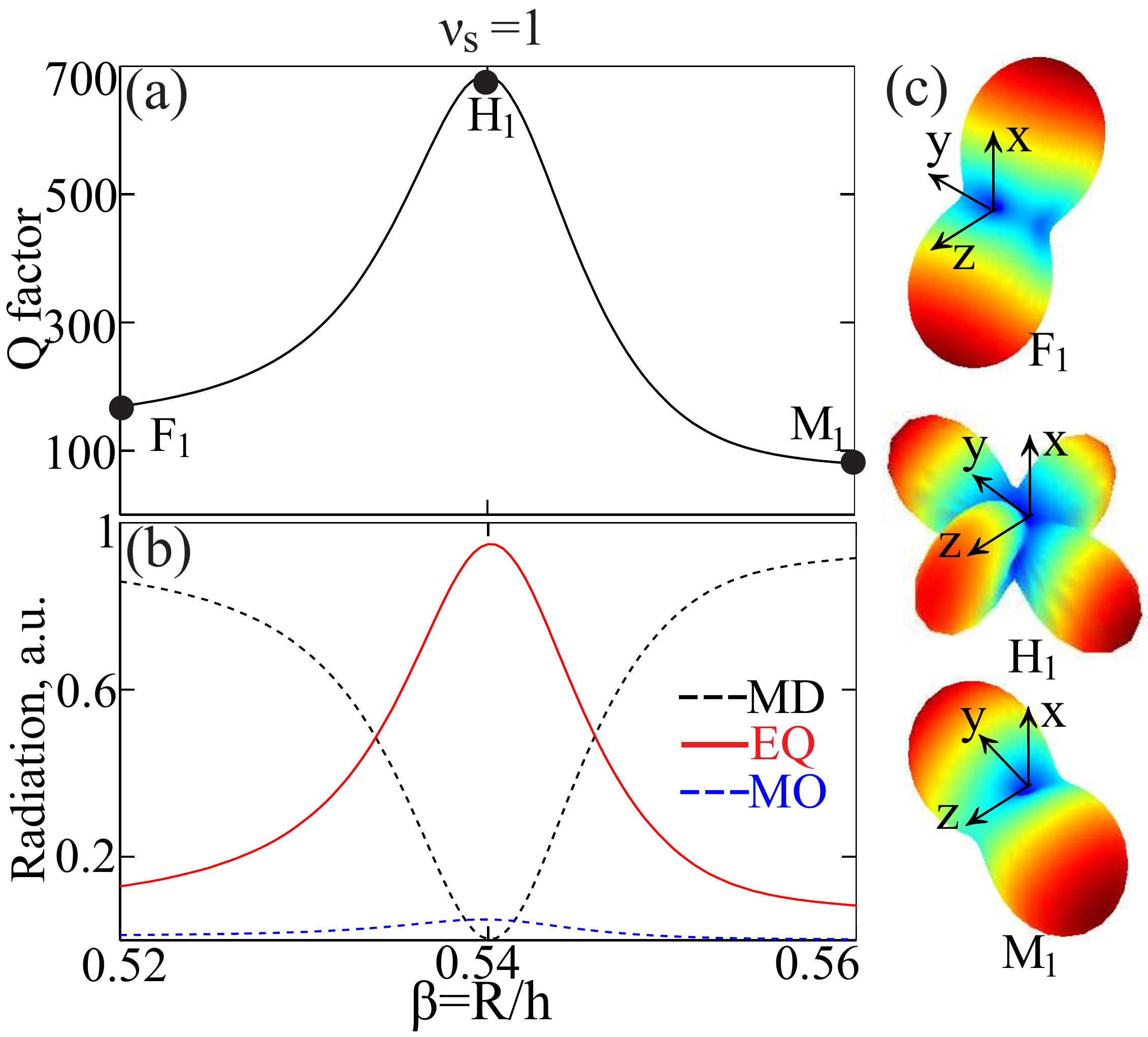}} \caption{\small (a) Evoluations of Q factors, and (b) the  radiated power from all multipoles that are not negligible, for the supermodes located on the left branch in the region marked in Fig.~\ref{fig1}(c), with $\nu_\textbf{s}=1$. In (a) three supermodes are indicated by points $\textbf{F}_1$, $\textbf{H}_1$, and $\textbf{M}_1$ (for which $\beta=0.52,~0.54,~0.56$, respectively) and the corresponding far-field radiation patterns are shown in (c).}\label{fig3}
\end{figure}

As is shown schematically in Fig.~\ref{fig1}(a), we start with this widely employed configuration~\cite{RYBIN_2017_Phys.Rev.Lett._High,BOGDANOV_2018_ArXiv180509265Phys._directa,BORDO_2010_Phys.Rev.B_Modela,ABUJETAS_2015_ACSPhotonics_Unraveling,LANDREMAN_2016_Opt.ExpressOE_FabryPerot,ABUJETAS_ACSPhotonics_highcontrast_2017-2,FROLOV_2017_NanoLett._NearField}
(refer to Ref.~\cite{Supplemental_Material_2} for investigations into other scenarios, in both strong and weak coupling regimes). The resonator is a nonmagnetic dielectric rod of height $h$, radius $R$, and relative permittivity $\varepsilon=40$ that is accessible for semiconductors~\cite{KIVSHAR_Opt.PhotonicsNews_metaoptics_2017-1}. Certain eigenmodes of interest supported by the resonator can be excited with a plane wave with wave-vector $\textbf{k}$ along $\textbf{z}$ and perpendicular to the rod axis $\textbf{y}$. The incident wave can be either $\textbf{\textit{p}}$-($\mathbf{E}||\mathbf{x}$) or $\textbf{\textit{s}}$-polarized ($\mathbf{E}||\mathbf{y}$). The plane wave scattering cross section spectra of this resonator with respect to normalized radius $\alpha=kR$ and aspect ratio $\beta=R/h$ are shown in Figs.~\ref{fig1}(b) and (c), for $\textbf{\textit{p}}$ and $\textbf{\textit{s}}$ polarizations, respectively~\cite{Supplemental_Material_2}. Altogether four representative anti-crossing regions of strong two-mode coupling are marked: within each region the two branches of supermodes, which are induced by the coupling between Mie-type and Fabry-Perot-type modes that share the same azimuthal quantum number $\nu$~\cite{RYBIN_2017_Phys.Rev.Lett._High,BOGDANOV_2018_ArXiv180509265Phys._directa,BORDO_2010_Phys.Rev.B_Modela,ABUJETAS_2015_ACSPhotonics_Unraveling,LANDREMAN_2016_Opt.ExpressOE_FabryPerot,
ABUJETAS_ACSPhotonics_highcontrast_2017-2},  are pinpointed by circles. The supermodes are quasi-normal modes~\cite{LALANNE__LaserPhotonicsRev._Light}, which can be characterized by complex eigenfrequencies $\widetilde\omega=\widetilde\omega_1+i\widetilde\omega_2$; the circles are centered at ($\widetilde\omega_1R/c$, $\beta$), where $c$ is the speed of light and the Q factor can be obtained through $Q=\widetilde\omega_1/(2\widetilde\omega_2)$~\cite{Supplemental_Material_2}. As will be further verified in Figs.~\ref{fig2} and \ref{fig3}, for $\textbf{\textit{p}}$ polarization, $\nu_\textbf{p}=0,1,2...$ corresponds to the magnetic dipole (MD), electric dipole (ED), electric quadrupole (EQ)..., while for $\textbf{\textit{s}}$ polarization, $\nu_\textbf{s}=0,1,2...$ corresponds to the electric dipole (ED), magnetic  dipole (MD), magnetic quadrupole (MQ)...~\cite{kallos2012resonance,LIU_Phys.Rev.Lett._generalized_2017}.

Next we examine the higher Q-factor supermodes, which locate on the higher-energy right (negative coupling) and lower energy left (positive coupling) branches for $\textbf{\textit{p}}$ and $\textbf{\textit{s}}$ polarizations respectively, with  mode cross coupling strengths of opposite signs~\cite{FRIEDRICH_Phys.Rev.A_interfering_1985-1,AWAI_2007_Electron.Commun.Jpn.PartIIElectron._Coupling}.  We investigate in detail the process of Q-factor enhancement for those modes, while noting that the supermodes on the other branches experience significant Q-factor suppression, as is shown in Ref.~\cite{Supplemental_Material_2}. In the following,  we categorize the high-Q supermodes into two groups of $\textbf{\textit{p}}$ and $\textbf{\textit{s}}$ polarizations. It basically means only that those supermodes can be excited by the normally incident plane waves of corresponding polarizations respectively. In conventional studies it is taken for granted that the Q-factor of a resonance is inextricably linked to the total radiation loss, while its relevance to radiation distributions along different directions is basically disregarded (see Ref.~\cite{Supplemental_Material_2} for  more general discussions based on the two mode coupling model, clarifying why the connection we reveal is generally hidden). In contrast, here through the approach of multipolar expansions~\cite{jackson1962classical,Bohren1983_book,DOICU_light_2006,POWELL_Phys.Rev.Applied_interference_2017,LIU_2018_Opt.Express_Generalized}, we manage to  reveal that the Q factor and radiation pattern are not segregated properties of resonances but rather they are subtlety connected with each other.

The radiated fields of each eigenmode can be fully expanded into two categories of multipoles of electric and magnetic natures, which correspond respectively to two sets of spherical harmonics with expansion coefficients of $a_{\rm {nm}}$ and $b_{\rm {nm}}$ (see Ref.~\cite{Supplemental_Material_2} for more information). To be more specific, the radiated power of each electric multipole and magnetic multipole of order $n$ is proportional respectively to ${\sum\nolimits_{m=-n}^{n} (2n+1){|{a_{\rm{nm}}}|} ^2}$ and ${\sum\nolimits_{m=-n}^{n} (2n+1){|{b_{\rm{nm}}}|} ^2}$, with $n=1,2,3$ corresponding to dipole, quadrupole, octupole, and so on and so forth. The expansion coefficients $a_{\rm {nm}}$ and $b_{\rm {nm}}$ can be calculated through either current integrations or  direct expansions of the radiated fields~\cite{jackson1962classical,Bohren1983_book,DOICU_light_2006}. It's worth mentioning that modes  and multipoles are two different concepts, though there are direct correspondences between them in highly symmetric structures such as spheres~\cite{jackson1962classical,Bohren1983_book,DOICU_light_2006,POWELL_Phys.Rev.Applied_interference_2017}. For most open photonic systems without symmetry, there is no such direct mapping between them and in stead each eigenmode can be expanded into a series of multipoles and not every multipole has an eigenmode correspondence~\cite{jackson1962classical,Bohren1983_book,DOICU_light_2006,POWELL_Phys.Rev.Applied_interference_2017}. For investigations related to far-field radiation properties, the language of multipoles is superior, as the eigenmode itself is not directly linked to angular radiation pattern~\cite{LIU_2018_Opt.Express_Generalized}.

\begin{figure}[tp]
\centerline{\includegraphics[width=8.8cm]{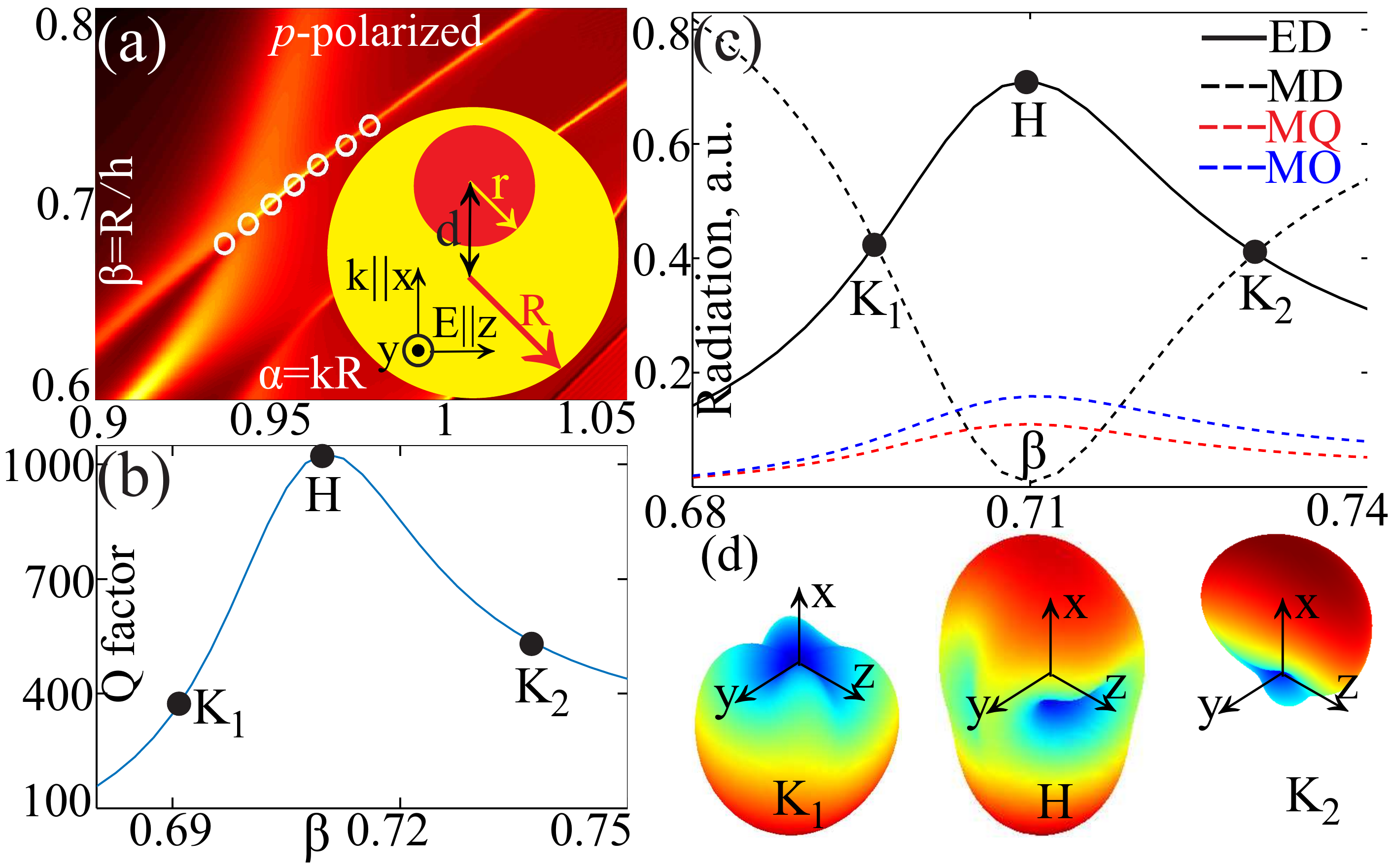}} \caption{\small (a) Scattering spectra of a $\textbf{\textit{p}}$-polarized exciting plane wave,  with a supermode branch indicated by circles. Inset: schematic of the asymmetric eccentrical core (air hole)-shell ($\varepsilon=40$) resonator with inner radius $r$, out radius $R$, center off-set $d$ and height $h$ (not shown). (b) Evolutions of Q factors and (c) the radiated power from all multipoles, with three supermodes indicated by points $\textbf{K}_1$, \textbf{H} and $\textbf{K}_2$ (for which $\beta=0.695,~0.7025,~0.73$, respectively). The corresponding radiation patterns at those points are shown in (d).}
\label{fig4}
\end{figure}

The results obtained for the supermodes that can be excited with the $\textbf{\textit{p}}$-polarized plane wave are summarized in Fig.~\ref{fig2}, with all the three right branches marked in Fig.~\ref{fig1}(b) investigated. We emphasize here that the eigenmodes of a resonator are determined solely by consisting materials and geometric configuration, while have nothing to do with the exciting source, being it plane wave, dipole emitter or source of any other forms. The plane wave scattering spectra shown in Fig.~\ref{fig1} provide a direct guide as to where the Q-factor enhanced supermodes locate, whereas we have to keep in mind that such a guide is incomplete, as no information of many other modes that cannot be excited by planes waves can be extracted from such spectra. For all three branches investigated, the Q factors of the supermodes are significantly enhanced in anti-crossing regions and for each case there is a optimum point (indicated by \textbf{H} point) where the Q factor reaches the maximum [see Figs.~\ref{fig2}(a)-(c)]. The corresponding results (radiated power from each contributing multipole with the total radiated power normalized, as is also the case in Figs.~\ref{fig3} and \ref{fig4}) obtained through multipolar expansions of supermodes are shown in Figs.~\ref{fig2}(d)-(f), where the radiated power of the multipoles that are not shown is negligible. It is clear that for all cases of different $\nu_\textbf{p}$, the Q-factor enhancement is intrinsically accompanied by the multipolar conversions from lower to higher orders: mainly MD to MO (magnetic octupole) for $\nu_\textbf{p}=0$ [Fig.~\ref{fig2}(d)], ED to MQ for $\nu_\textbf{p}=1$ [Fig.~\ref{fig2}(e)], and EQ to MO for $\nu_\textbf{p}=2$ [Fig.~\ref{fig2}(f)]. Moreover, the positions of maximum Q factors always coincides with the peaks of the dominating higher-order multipoles, and away from those optimum points, multipoles are reversely converted from higher to lower orders with decreasing Q factors. Moreover, at the optimum points, the higher orders the dominating multipole are of, the higher Q factors can be achieved.

To further verify the results from multipolar expansions, along each supermode branch we have selected three supermodes indicated by $\textbf{M}_{0,1,2}$, $\textbf{H}_{0,1,2}$, and $\textbf{F}_{0,1,2}$ (the subscript indicates the value of $\nu$, as is also the case in Fig.~\ref{fig3}) in Figs.~\ref{fig2}(a)-(c). This includes not only highest Q-factor modes $\textbf{H}_{0,1,2}$, but also lower Q-factor modes $\textbf{M}_{0,1,2}$ and  $\textbf{F}_{0,1,2}$ that bear more resemblances respectively to Mie-type and Fabry-Perot-type modes, due to weak coupling between them at regions relatively far from the optimum points~\cite{RYBIN_2017_Phys.Rev.Lett._High,BOGDANOV_2018_ArXiv180509265Phys._directa}. The corresponding far-field radiation patterns of those supermodes are shown correspondingly in Figs.~\ref{fig2}(g)-(i) [also refer to Ref.~\cite{Supplemental_Material_2} for the corresponding near-field distributions of those supermodes], which agree well with what is shown in Figs.~\ref{fig2}(d)-(f).  At $\textbf{H}_{0,1,2}$ points there are more radiation lobes (the lobe number is proportional to the multipole order) than at other two points, confirming the conversions between multipoles of different orders.

We proceed to discuss the supermodes that can be excited with the $\textbf{\textit{s}}$-polarized plane wave, as is marked in Fig.~\ref{fig1}(c) with $\nu_\textbf{s}=1$. The results of Q-factor evolutions, multipolar radiation spectra, and the corresponding far-field radiation patterns of the selected supermodes are shown in Figs.~\ref{fig3}(a)-(c), respectively. The same conclusion with respect to the connection between Q-factor enhancement and multipolar conversions can be drawn. We note here that: on one hand, the multipolar conversion mechanism can be widely employed to enhance Q factors, not only for individual particles as we show here, but also for more sophisticated photonic crystal cavities~\cite{JOHNSON_2001_Appl.Phys.Lett._Multipolecancellationa,KARALIS_2004_Opt.Lett._Discretemode} and probably many other sorts of resonators~\cite{LEPETIT_Phys.Rev.B_controlling_2014,BULGAKOV_2018_ArXiv180406626Phys._Fibers}; on the other hand however, we have to keep in mind that not all sorts of Q-factor enhancements can be attributed to such a mechanism, such as those obtained relying on resonators  consisting of materials of unusual effective parameters where there is no effective mode coupling effect~\cite{SILVEIRINHA_Phys.Rev.A_trapping_2014,MONTICONE_Phys.Rev.Lett._embedded_2014,LIU_Phys.Rev.B_qfactor_2016} (refer also to Ref.~\cite{Supplemental_Material_2}, where it is shown that the left-branch supermodes marked in Fig.~\ref{fig1}(b) and  the right-branch supermodes marked in Fig.~\ref{fig1}(c) experience significant Q-factor suppressions without multipolar conversions).

Here we have gained a deeper insight into the fundamental two-mode coupling model, by successfully bridging the two features of Q-factor enhancement and multipolar conversions from lower to higher orders. Now the question is: what can we do with it?  Up to now, we have already obtained supermodes of simultaneous high-Q factors and subwavelength mode volumes with symmetric dielectric rods. Besides high-Q factors and subwavelength mode volumes,  for many applications relying on nanoscale elements such as low-threshold lasers, antennas, sensors and single-photon sources~\cite{vahala_optical_2004,VUKOBRATOVICH_2018__Fundamentalsa}, modes with asymmetrically directional radiation patterns are even more desired. This leads to a fundamental but also technologically interesting question: is it feasible to synchronize all three desirable features of sub-wavelength mode volume, high-Q factor and unidirectional far-field radiation within one nanoscale resonator?

Our next goal is to add the third desired feature to the supermodes achieved to make their radiations unidirectional. We term such supermodes as \textit{Kerker supermodes}, as according to the generalized Kerker theory the unidirectional radiations should consist of a series of multipoles of different natures and/or different orders~\cite{LIU_2018_Opt.Express_Generalized,Kerker1983_JOSA,jahani_alldielectric_2016,KUZNETSOV_Science_optically_2016}. Though it is shown in lots of previous studies that many symmetric structures may exhibit highly directional scattering (radiation) patterns, it should be noted that such directionality always comes from the co-excitations and interferences of several eigenmodes, and thus it is generically dependent on external sources~\cite{LIU_2018_Opt.Express_Generalized,KUZNETSOV_Science_optically_2016}. To obtain an eigenmodes with intrinsic directional radiation that is independent on the exciting source, the basic requirement is to break the rotational symmetry of the resonator~\cite{cao2015_RMP_dielectric,LEE_2002_Phys.Rev.Lett._Observation,WIERSIG_2006_Phys.Rev.A_Unidirectional,SONG_2010_Phys.Rev.Lett._Directionala}. Our further explorations for Kerker supermodes are still based on the dielectric rod shown in Fig.~\ref{fig1}(a), the symmetry of which is broken by introducing an extra eccentric cylindrical air hole of radius $r=0.25$R and center-to-center displacement $d=0.675$R along the $\textbf{x}$ direction, as is shown in the inset of Fig.~\ref{fig4}(a). Following our previous approach, we also show in Fig.~\ref{fig4}(a) the scattering spectra with a $\textbf{\textit{p}}$-polarized incident plane wave ($\textbf{k}||\textbf{x}$ and $\textbf{E}||\textbf{z}$), from which the supermodes with enhanced Q factors can be identified. We have further pinpointed by circles one supermode branch in the scattering spectra, for which the dependence of Q factors and multipolar radiation spectra on aspect ratios are shown respectively in Figs.~\ref{fig4}(b) and (c).

For this asymmetric resonator however, there is no well defined azimuthal quantum number to characterize such supermode branch, nor is it easy to identify which sets of modes couple and hybridize with one another to induce it. Nevertheless, almost identical to the results shown in  Figs.~\ref{fig2} and \ref{fig3}, for the asymmetric case the Q-factor enhancement is also accompanied by multipolar conversions [see Figs. \ref{fig4}(b) and (c)].  Along the supermode branch we have pinpointed three points [the corresponding far-field radiation patterns are shown in Fig. \ref{fig4}(d)]. The highest Q factor is obtained at point \textbf{H} and the radiation power from ED and MD equal to each other at points $\textbf{K}_1$ and $\textbf{K}_2$. At point \textbf{H}, the coexistence of ED, MQ, MO, and interference among them can only render the radiation slightly asymmetric along $x$ direction [see Fig. \ref{fig4}(d)], since here the contribution from ED is still dominant over other multipoles. At points $\textbf{K}_1$ and $\textbf{K}_2$, the interference among the multipoles excited (mainly ED and MD) renders the radiation highly asymmetric [see Fig. \ref{fig4}(d)] (refer to Ref.~\cite{Supplemental_Material_2} for the corresponding near-field distributions of those supermodes, which are also asymmetric due to geometric symmetry breaking). These Kerker supermodes share the same mechanism as that of the Kerker scattering of the first and second kinds, with suppressed forward or backward scatterings that originate from destructive interferences between EDs and MDs of close magnitudes~\cite{LIU_2018_Opt.Express_Generalized,Kerker1983_JOSA,jahani_alldielectric_2016,KUZNETSOV_Science_optically_2016}.

In conclusion, we have discovered a hidden dimension of the fundamental two-mode coupling model, revealing the subtle connection between Q-factor enhancement and redistributions of far-field angular radiations that originate from multipolar conversions from lower to higher orders~\cite{Supplemental_Material_2}. Based on this discovery and the generalized Kerker effect that can result in radiation directionality, we have achieved with asymmetric nanoscale dielectric resonators the Kerker supermodes with simultaneously  high-Q factors, subwavelength made volumes, and unidirectional radiation patterns. Here we have confined our discussions of this fundamental model within passive and nonmagnetic photonic systems, and it is expected that a natural extension to more sophisticated configurations (periodic, quasi-periodic or random that involve many-mode coupling) with active materials (with gain, chirality or nonlinearity~\cite{Supplemental_Material_2,FENG_2017_Nat.Photonics_NonHermitian,WAGNIERE_2007__Chirality,CARLETTI_2018_Phys.Rev.Lett._Giant}) would render much broader freedom for manipulations of strong light-matter interactions at the nanoscale.  We believe that the neglected new dimension we have revealed here can spur many optical resonator related fundamental explorations of cavity electrodynamics and incubate a wide range of applications related to nanoscale lasers, sensors, single-photon sources and many other optical elements. Our discovery can also shed new light on lots of other interdisciplinary fields (such as optomechanics, exciton-polaritonics, phonon-polaritonics and so on) that involve interactions and hybridizations among modes of mechanical, phononic, electronic or other hybrid natures.

We thank Q. H. Song, David Powell and A. E. Miroshnichenko for useful discussions, and acknowledge the financial support from National Natural Science
Foundation of China (Grant No. 61405067, 11174250, 11874026, 11404403 and 11874426), and the Outstanding Young Researcher Scheme of National University of Defense Technology.
%

\newpage

\newpage

\setcounter{figure}{0}
\makeatletter
\renewcommand{\thefigure}{S\@arabic\c@figure}
\makeatother

\makeatletter
\renewcommand{\theequation}{S\@arabic\c@equation}
\makeatother

\textbf{Supplemental Material for ``Multipolar Conversion Induced Subwavelength High-Q Supermodes with Unidirectional Radiations"}

\section{\uppercase\expandafter{\romannumeral1}. Plane wave scattering spectra, eigenfrequencies, radiation patterns and Q factors of the eigenmodes.}

The plane wave ($\textbf{\textit{p}}$-polarized or $\textbf{\textit{s}}$-polarized) scattering cross section spectra shown in Figs.~1(b) and (c) and Fig.~4(a) are obtained with commercial software CST Microwave Studio (https://www.cst.com/). For the finite dielectric resonators (both symmetric and asymmetric ones) studied in this work, we employ a commercial software package COMSOL MULTIPHYSICS (https://www.comsol.com) to calculate the eigenfrequencies of the eigenmodes in Figs.~1 \& 4, and Figs.~\ref{figr2} \& ~\ref{figr3} (shown below), and adopt the method described in Ref.~\cite{YAN_2018_Phys.Rev.B_Rigorous} for calculation when lossy and dispersive materials are involved (see Fig.~\ref{figr4} below). Since finite resonators are open systems, the eigenmodes are quasi-normal modes and characterized by complex eigenfrequencies with both noneligible real and imaginary parts: $\widetilde\omega=\widetilde\omega_1+i\widetilde\omega_2$. For all eigenmodes, the corresponding far-field radiation patterns (shown in Figs.~2-4, and Figs.~\ref{figs3} and ~\ref{figs4}), near-field distributions (shown in Figs.~\ref{figs1}-\ref{figs4}), and eigenfrequencies $\widetilde\omega_{1,2}$  can be directly calculated using COMSOL. Then the Q factors of those eigenmodes can be obtained with the following relation~\cite{HAUS_1983__Waves,Yariv2006_book_photonics}:
\begin{equation}
\label{Q-factor}
 Q=\frac{\widetilde\omega_1}{2\widetilde\omega_2}.
\end{equation}
\section{\uppercase\expandafter{\romannumeral2}.  Multipolar expansions for radiations of eigenmodes.}
The radiated fields of each eigenmode can be expanded as~\cite{jackson1962classical,Bohren1983_book,DOICU_light_2006}:
\begin{equation}
\label{multipolar expansion}
{\textbf{E}_{\rm{rad}}} = \sum\limits_{n = 1}^\infty \sum\limits_{m =  - n}^n {{a_{nm}}\textbf{N}_{nm}^{(3)}(r) + {b_{nm}}\textbf{M}_{nm}^{(3)}},
\end{equation}
where $\textbf{N}_{nm}$ and $\textbf{M}_{nm}$ are vector spherical harmonics, the expressions of which can be found in Refs.~\cite{jackson1962classical,Bohren1983_book,DOICU_light_2006}; $\textbf{M}_{nm}$ is tangential to the radial direction $\textbf{r}\cdot\textbf{M}_{nm}=0$ and $\textbf{N}_{nm}$ is proportional to $\nabla\times\textbf{M}_{nm}$; $\textbf{N}_{nm}$ (and thus $a_{nm}$) corresponds to the electric multipole of order $n$ and  $\textbf{M}_{nm}$ (and thus $b_{nm}$) corresponds to the magnetic multipole of order $n$. According to Eq.~(\ref{multipolar expansion}), with the  field components associated with each eigenmode ${\textbf{E}_{\rm{rad}}}$ obtained (which can be calculated with COMSOL), $a_{nm}$ and $b_{nm}$ can be achieved through:
\begin{figure}
\centerline{\includegraphics[width=7.5cm]{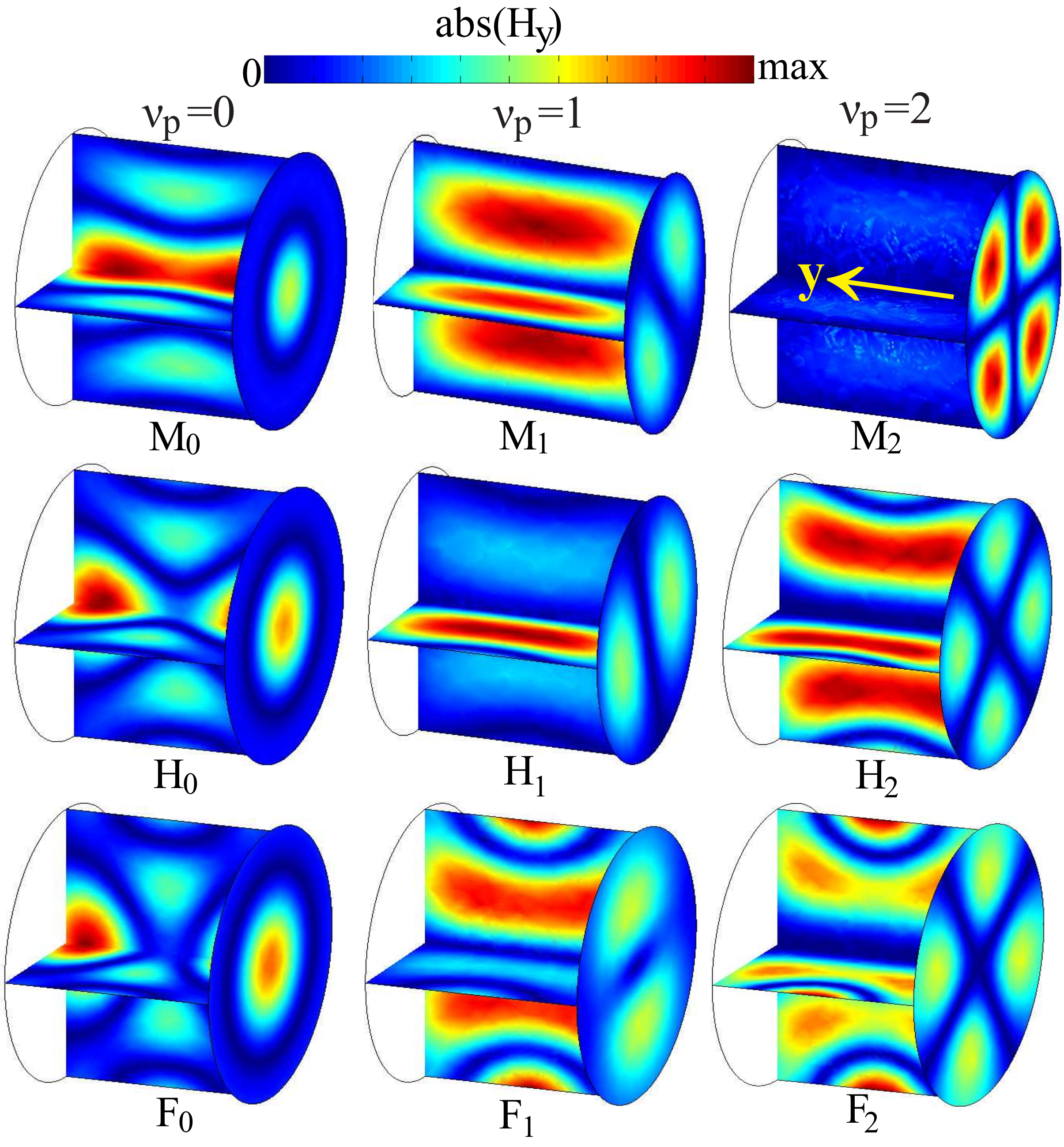}} \caption{\small Near-field distributions in terms of axial magnetic field magniude abs($\textbf{H}_y$) are shown for the nine eigenmodes indicated by black dots in Fig.~2. Both the fields on the $\textbf{y}$-perpendicular plane and other two $\textbf{y}$-parallel planes are shown. }\label{figs1}
\end{figure}
%

\begin{equation}
\label{coefficients}
\begin{split}
{a_{nm}} = {{\int\!\!\!\int\limits_\textbf{S} {{{\bf{E}}_{\rm{rad}}} \cdot {{\bf{N}}^{(3)}_{nm}}^*d\textbf{s}} } \over {{}\int\!\!\!\int\limits_\textbf{S} {{{\bf{N}}^{(3)}_{nm}} \cdot {{\bf{N}}^{(3)}_{nm}}^*d\textbf{s}} }},\\
{b_{nm}} = {{\int\!\!\!\int\limits_\textbf{S} {{{\bf{E}}_{\rm{rad}}} \cdot {{\bf{M}}^{(3)}_{nm}}^*d\textbf{s}} } \over {{}\int\!\!\!\int\limits_\textbf{S} {{{\bf{M}}^{(3)}_{nm}} \cdot {{\bf{M}}^{(3)}_{nm}}^*d\textbf{s}} }},
\end{split}
\end{equation}
where $\ast$ denotes complex conjugation and the integration is implemented over a close spherical surface $\textbf{S}$ that encloses the rod resonator. We have implemented a finite element scheme to efficiently carry out the numerical integrations with high precision. The idea is to employ the Lebedev quadrature rule to discrete the spherical surface into triangles on which a simple quadrature rule is implemented, and thus $a_{nm}$ and $b_{nm}$ in Eqs.~(\ref{multipolar expansion}) and (\ref{coefficients}) are evaluated. Then the radiated power of each eigenmode can be decomposed into those contributed by electric multipoles of different orders $n$ that are proportional to ${\sum\nolimits_{m=-n}^{n} (2n+1){|{a_{\rm{nm}}}|} ^2}$ and to those contributed by magnetic multipoles of different orders $n$ that are proportional to ${\sum\nolimits_{m=-n}^{n} (2n+1) {|{b_{\rm{nm}}}|} ^2}$. In our study reported in this work about 3D structures, only dipoles, quadrupoles and octupoles are involved, which correspond to $n=1, 2, 3$, respectively. To be more specific, for example, the radiated power of the electric dipole is proportional to ${\sum\nolimits_{m=-1}^{1} 3{|{a_{\rm{1m}}}|} ^2}$ and that of the magnetic quadrupole is proportional to ${\sum\nolimits_{m=-2}^{2} 5{|{b_{\rm{2m}}}|} ^2}$, and so on and so forth. For the multipolar radiation spectra shown in Figs.~2-4 and Figs.~\ref{figs3} and \ref{figs4}, the total radiated power of each eigenmode is normalized.

The discussions above on multipolar expansions are for three dimensional (3D) cases, the expressions for which can be easily extended to two dimensional (2D) cases. The eigenmodes of 2D structures can be categorized as transverse electric (TE, electric fields on plane and the magnetic fields along the out-of-plane $\textbf{z}$ direction) or transverse magnetic (TM, magnetic fields on plane and the electric fields along the out-of-plane $\textbf{z}$ direction).  The radiated fields of each eigenmode can be expanded respectively as~\cite{jackson1962classical,Bohren1983_book,DOICU_light_2006}:
\begin{equation}
\label{multipolar expansion_cylinder}
\begin{split}
\textbf{E}_{\rm{rad}}^{\rm {\textbf{TE}}} = \sum\limits_{m =  - \infty}^{\infty} {{a_{m}}\textbf{N}_{m}^{(3)}(r)},\\
\textbf{E}_{\rm{rad}}^{\rm {\textbf{TM}}} = \sum\limits_{m =  - \infty}^{\infty} {{b_{m}}\textbf{M}_{m}^{(3)}(r)},
\end{split}
\end{equation}
where $\textbf{N}_{m}$ and $\textbf{M}_{m}$ are vector cylinderical harmonics, the expressions of which can be found in Refs.~\cite{jackson1962classical,Bohren1983_book,DOICU_light_2006}; $\textbf{M}_{m}$ is parallel to $\textbf{z}$ and $\textbf{N}_{m}$ is perpendicular to $\textbf{z}$.  For TE modes: $a_0$ corresponds to the magnetic dipole (MD); $a_{\pm 1}$ corresponds to the electric dipole (ED); $a_{\pm 2}$ corresponds to the electric quadrupole(EQ) and so on and so forth; while for TM modes: $b_0$ corresponds to the electric dipole (ED); $b_{\pm 1}$ corresponds to the magnetic dipole (MD); $b_{\pm 2}$ corresponds to the magnetic quadrupole(EQ) and so on and so forth~\cite{LIU_Phys.Rev.A_superscattering_2017}.  Similar to the 3D case, $a_{m}$ and $b_{m}$ can be obtained through:
\begin{equation}
\label{coefficients_cylinder}
\begin{split}
{a_{m}} = {{\!\!\!\int\limits_{\phi=0}^{2\pi} {{{\textbf{E}_{\rm{rad}}^{\rm {\textbf{TE}}}}} \cdot {{\bf{N}}^{(3)}_{m}}^*d\phi} } \over {{}\!\!\!\int\limits_{\phi=0}^{2\pi} {{{\bf{N}}^{(3)}_{m}} \cdot {{\bf{N}}^{(3)}_{m}}^*d{\phi}} }},\\
{b_{m}} = {{\!\!\!\int\limits_{\phi=0}^{2\pi} {{{\textbf{E}_{\rm{rad}}^{\rm {\textbf{TM}}}}} \cdot {{\bf{M}}^{(3)}_{m}}^*d\phi} } \over {{}\!\!\!\int\limits_{\phi=0}^{2\pi} {{{\bf{M}}^{(3)}_{m}} \cdot {{\bf{M}}^{(3)}_{m}}^*d{\phi}} }}.
\end{split}
\end{equation}
 Then the radiated power of each eigenmode can be decomposed into those contributed by multipoles of different orders $n=|m|$, which is proportional to ($|a_n^2|+|a_{-n}^2|$) ($|b_n^2|+|b_{-n}^2|$) for TE and TM cases, respectively. Without losing generality, the total radiated power of each eigenmode is normalized.

\section{\uppercase\expandafter{\romannumeral3}.   Near-field distributions of the eigenmnodes indicated in Figs.~2 and 4.}

\begin{figure}
\centerline{\includegraphics[width=8cm]{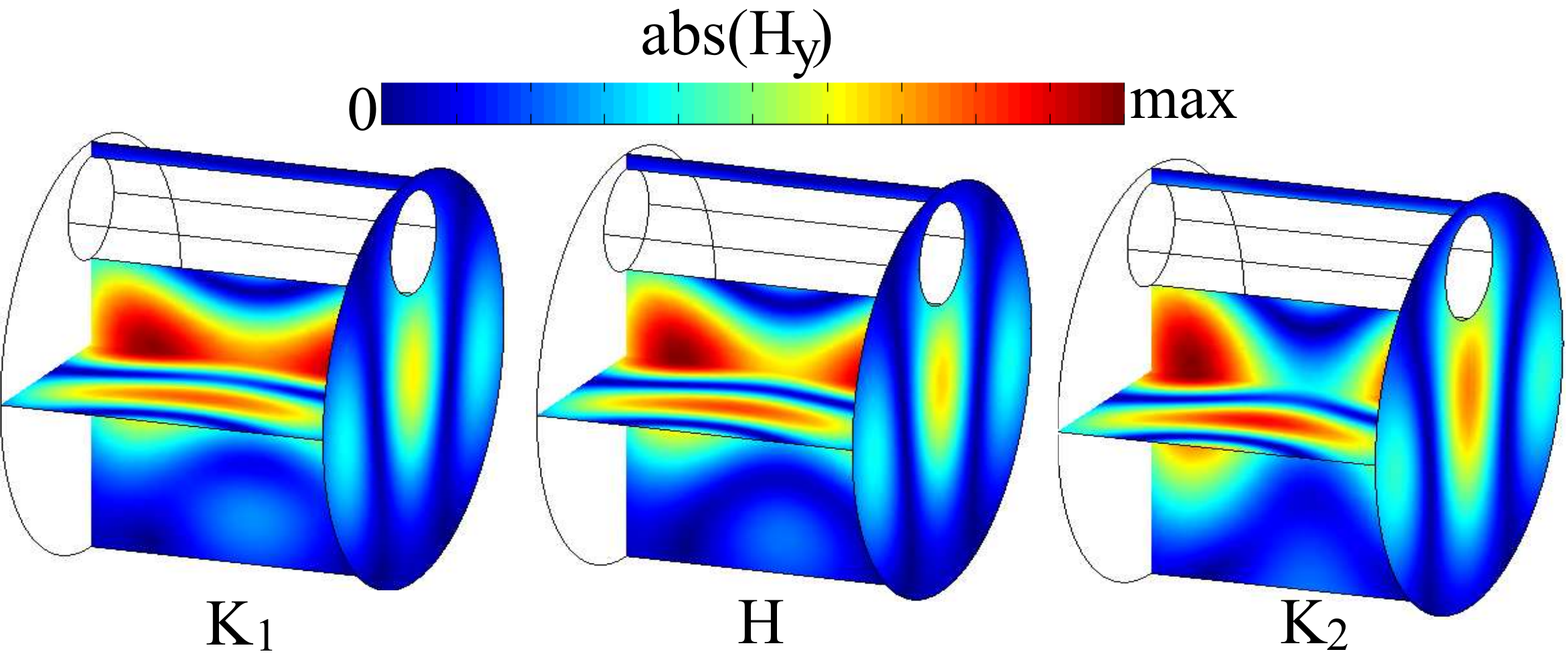}} \caption{\small Near-field distributions in terms of axial magnetic electric field magniude abs($\textbf{H}_y$) are shown for the three eigenmodes indicated by black dots in Fig.~4. Both the fields on the $\textbf{y}$-perpendicular plane and other two $\textbf{y}$-parallel planes are shown.}\label{figs2}
\end{figure}

\begin{figure}
\centerline{\includegraphics[width=8.8cm]{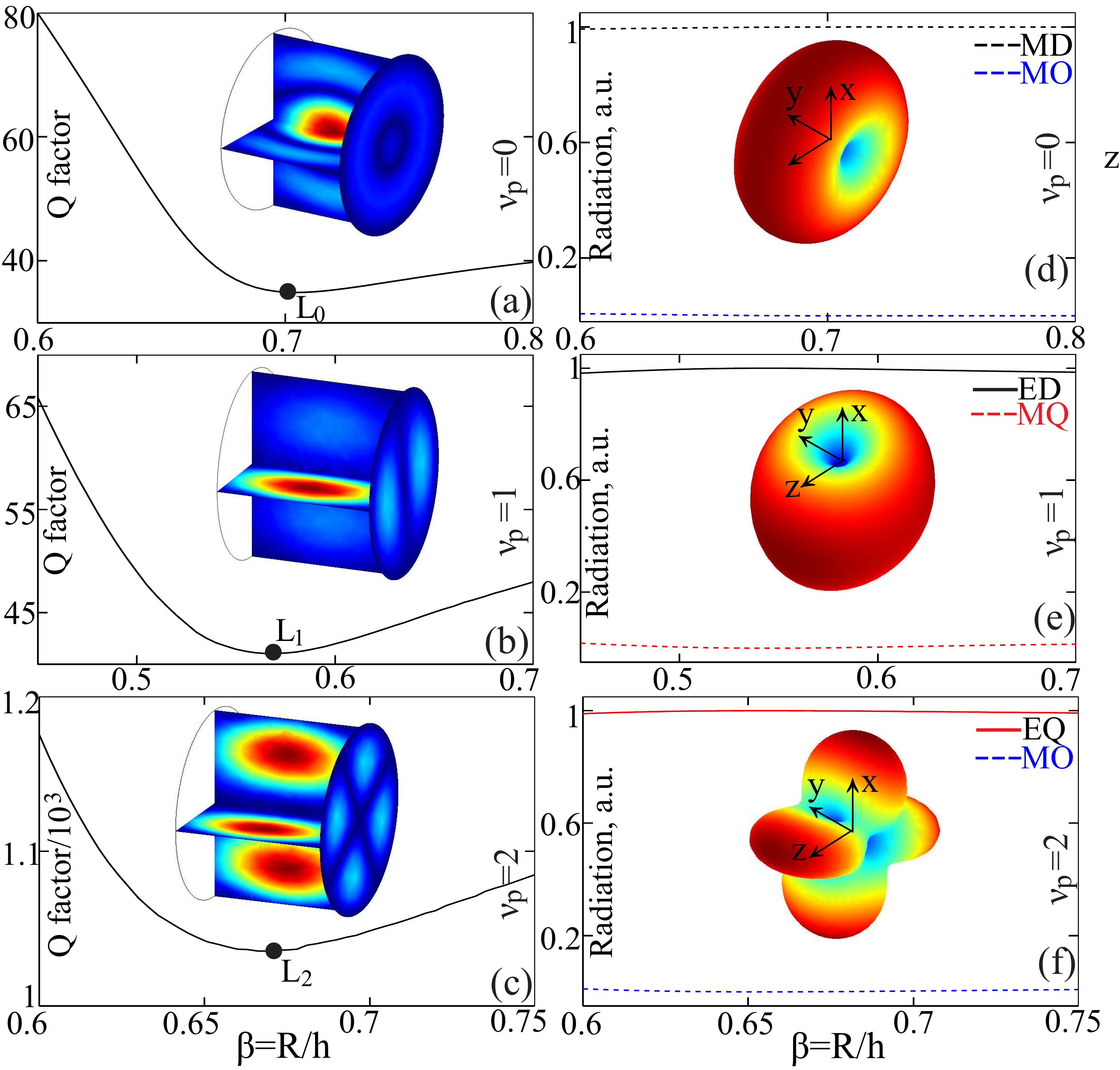}} \caption{\small (a)-(c) Evoluations of Q factors, and (d)-(f) the radiated power from all multipoles that are not negligible, for the supermodes located on the left branches in the three anti-crossing regions marked in Fig.~1(b), with $\nu_\textbf{p}=0,1,2$ respectively. For each case in (a)-(c), the supermodes with lowest Q factors are indicated by points $\textbf{L}_{0,1,2}$ (for which $\beta=0.704,~0.565,~0.669$, respectively), and the corresponding near-field distributions [abs($\textbf{H}_y$)] and far field radiation patterns are shown as insets in (a)-(f).}\label{figs3}
\end{figure}

To further verify the results of multipolar conversions that accompany the Q-factor enhancement shown in Fig.~2, we show in Fig.~\ref{figs1} the evolutions of the near fields [magnitude of axial magnetic electric field, abs($\textbf{H}_y$)] for the nine eigenmodes indicated. It is clear from Fig.~\ref{figs1} that the azimuthal quantum number $\nu_\textbf{p}$ agrees very well with the near-field distributions of the eigenmodes on the $\textbf{y}$-perpendicular plane: the number of side field lobes is $2\nu_\textbf{p}$~\cite{LIU_Phys.Rev.A_superscattering_2017}. Also as expected, during the multipolar conversion processes there exist significant changes of the near-field distributions. For the eigenmodes indicated in Fig.~3 that can be excited with $\textbf{\textit{s}}$-polarized plane waves, similar effects can be observed. For the eigenmodes indicated in Fig.~4 that are supported by dielectric resonator with broken symmetry, the near-field distributions are summarized in Fig.~\ref{figs2}. Compared to those field distributions shown in Fig.~\ref{figs1} that are always symmetric as required by the symmetry of the resonator, the  resonator with broken symmetry can support asymmetric eigenmodes for which unidirectional radiations are accessible.

\section{\uppercase\expandafter{\romannumeral4}. Q factor evolutions and multipolar radiation spectra for the left-branch supermodes marked in Fig.1(b) and right-branch supermodes marked in Fig.1(c).}

In Fig.~1(b) we have marked three anti-crossing regions of strong mode coupling. In the following studies we focus only on the right-branch supermodes with significant Q-factor enhancement [see Fig.~2].  In contrast, the supermodes on the left branches are expected to experience opposite Q-factor suppression processes in the anti-crossing regions~\cite{FRIEDRICH_Phys.Rev.A_interfering_1985-1,HEISS_Phys.Rev.E_repulsion_2000-1,RYBIN_2017_Phys.Rev.Lett._High,BOGDANOV_2018_ArXiv180509265Phys._directa}. The results obtained for supermodes that can be excited with $\textbf{\textit{p}}$-polarized plane wave are summarized in Fig.~\ref{figs3}, with all the three left branches circled in Fig.~1(b) investigated. For all three cases, the Q factors of the supermodes are significantly suppressed in the mode anti-crossing regions and there are optimum points (indicated by $\textbf{L}_{0,1,2}$ points) where the Q factors reaches their minimums [see Figs.~\ref{figs3}(a)-(c)]. The corresponding results (radiated power from each contributing multipole, with the total radiated power of the supermode normalized) obtained through multipolar expansions of supermodes are shown in Figs.~\ref{figs3}(d)-(f), where the radiated power of the multipoles that are not shown is negligible. It's clear from Figs.~2(a)-(c) and Figs.~\ref{figs3}(a)-(c) that the positions of minimum Q factors (left branches) coincides well with the maximum Q-factor points for the supermodes on the corresponding right branches. At the same time, it is worth mentioning that: (i) for the right-branch supermodes, the Q-factor enhancement is induced by multipolar conversions from lower to higher orders; (ii) while for the left-branch supermodes, the Q-factor suppression is not accompanied by multipolar conversions from higher to lower orders, but rather the multipolar ratios are approximately fixed. We have also shown the near-field distributions and far-field radiation patterns of the supermodes with lowest Q factors as the insets of Figs.~\ref{figs3}(a)-(f).

\begin{figure}
\centerline{\includegraphics[width=8.8cm]{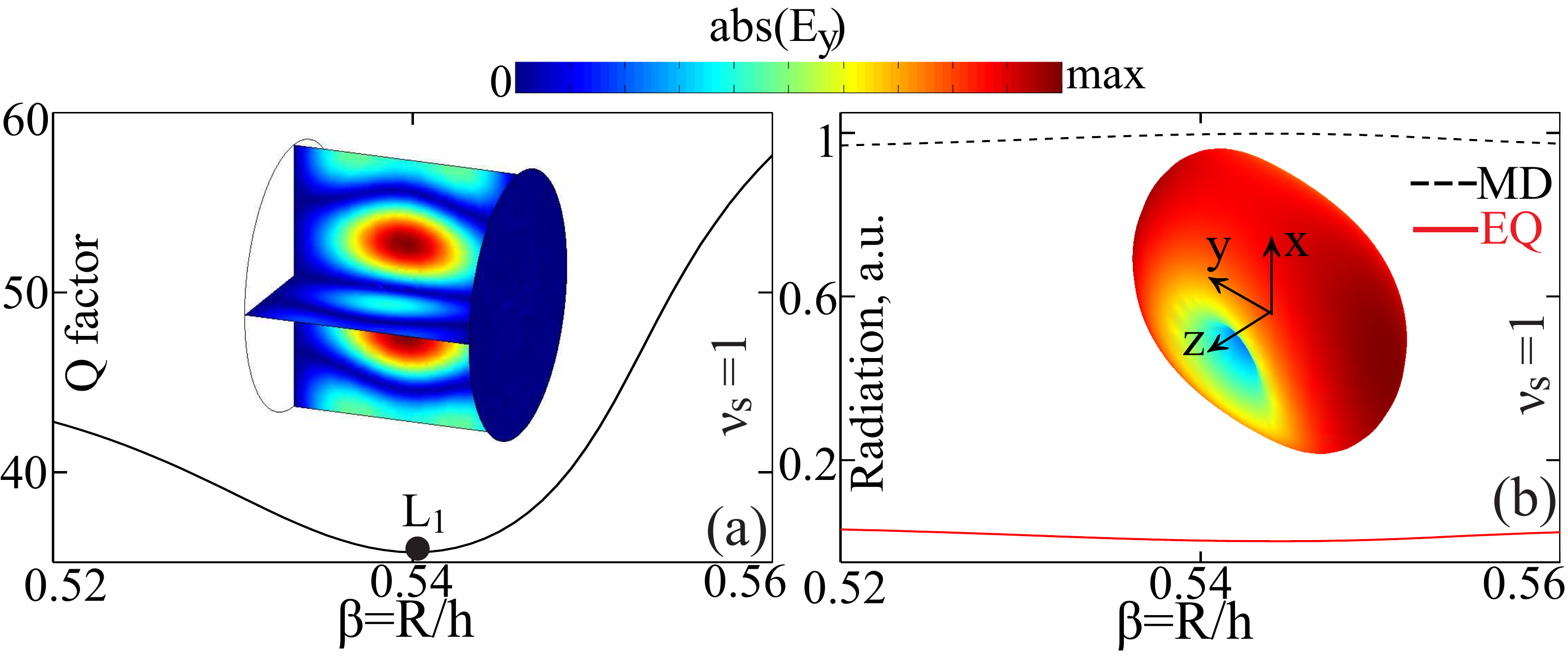}} \caption{\small (a) Evoluations of Q factors, and (b) the radiated power from all multipoles that are not negligible, for the supermodes located on the right branch in the anti-crossing region marked in Fig.~1(c), with $\nu_\textbf{s}=1$. The supermode with lowest Q factor is indicated by  $\textbf{L}_{1}$ ($\beta=0.54$), and the corresponding near-field distribution [abs($\textbf{E}_y$)] and far field radiation pattern are shown as insets in (a) and (b).}\label{figs4}
\end{figure}

In Fig.~1(c) however, the supermodes on the left branch experience Q-factor enhancement [see Figs.~3(a) and (b)] while those on the right branch experience Q-factor suppression [see Fig.~\ref{figs4}(a)]. This is sharply different from those anti-crossing regions marked in Fig.~1(b) [see Figs.~2(a) and (b)], which basically indicates mode cross coupling strengths of opposite signs, corresponding to positive coupling ($\textbf{\textit{s}}$ polarization) and negative coupling ($\textbf{\textit{p}}$ polarization), respectively~\cite{FRIEDRICH_Phys.Rev.A_interfering_1985-1}. The results obtained for supermodes that can be excited with $\textbf{\textit{s}}$-polarized plane wave are summarized in Fig.~\ref{figs4}, with the right-branch supermodes circled in Fig.~1(c) investigated. The Q factors of the supermodes are significantly suppressed in the anti-crossing regions and there is an optimum point (indicated by $\textbf{L}_{1}$) where the Q factor reaches its minimum [see Fig.~\ref{figs4}(a)]. The corresponding results (radiated power from each contributing multipole, with the total radiated power of the supermode normalized) obtained through multipolar expansions of supermodes are shown in Fig.~\ref{figs4}(b), where the radiated power of the multipoles that are not shown is relatively negligible. It's clear from Fig.~3(a) and Fig.~\ref{figs4}(a) that the position of minimum Q factor (right branch) coincides with the maximum Q-factor point for the supermode on the corresponding left branch. Similar to the case of $\textbf{\textit{p}}$ polarization, the Q-factor suppression is not accompanied by multipolar conversions from higher to lower orders, but rather the multipolar ratios are approximately fixed. We have also shown the near-field distribution and far-field radiation pattern of the supermode with lowest Q factor as the insets of Figs.~\ref{figs4}(a)-(b).

\section{\uppercase\expandafter{\romannumeral5}. General discussions based on two mode coupling model and more scenarios with multipolar conversion induced Q-factor enhancement in both strong and weak coupling regimes.}

In this section, as a first step, based on the two mode coupling model we reveal how the conventional kind of scalar treatment (simply assign a wave function to the involved mode without caring much about the details of the vector field distribution) obscures the subtle connection we have discovered between Q-factor enhancement and multipolar conversions from lower to higher orders. As a next step, we try to attribute such a connection to hybridizations of modes of intrinsic vector natures, and also discuss the limitation of such an attribution. Most importantly,  at the end we provide evidence with sharply different configurations (both single and coupled resonators) to confirm that multipolar conversion induced Q-factor enhancement is more generic than being only valid for specific geometries, which is present in both strong and weak coupling regimes.

$\blacksquare$~~~\textsc{\textbf{Two mode coupling model and the Hidden connection}}\\
The mechanism we have revealed applies to the coupling between  quasi-bound (quasi-normal) modes, which themselves are also coupled to the continuum and can thus be characterized by complex eigenfrequencies. Coupling between those modes (indexed by $\alpha$ and $\beta$) can be described by the following interaction matrix~\cite{HEISS_Phys.Rev.E_repulsion_2000-1,WIERSIG_Phys.Rev.Lett._formation_2006-1}:
\begin{equation}
\label{matrix}
\rm \textbf{H}=
\begin{pmatrix}
 {{{\widetilde \omega }_\alpha }} & {{{\rm{\widetilde C}}_{\alpha \beta }}}  \cr
 {{{\rm{\widetilde C}}_{\beta \alpha }}} & {{{\widetilde \omega }_\beta }}  \cr
\end{pmatrix}
\end{equation}
where $\widetilde \omega_{\alpha,\beta}$ are the corresponding complex eigenfrequencies for modes $\alpha$ and $\beta$, and the off-diagonal terms are the coupling coefficients, which are generally complex for nonconservative open systems. For such an interaction matrix shown in Eq.~(\ref{matrix}), the eigenfrequencies for the supermodes of the coupled system can be expressed as:
\begin{equation}
\label{eigenfrequency}
{\widetilde \omega _ \pm }{\rm{ = }}{1 \over 2}\left( {{{{{\widetilde \omega }_\alpha }{\rm{ + }}{{\widetilde \omega }_\beta }}} \pm \sqrt {{{\left( {{{\widetilde \omega }_\alpha }{\rm{ - }}{{\widetilde \omega }_\beta }} \right)}^2} + 4{{\rm{C}}_{\alpha \beta }}{{\rm{C}}_{\beta \alpha }}} } \right),
\end{equation}
with the associated eigenvectors being:
\begin{equation}
\label{eigenvector}
{1 \over {2{{\rm{C}}_{\beta \alpha }}}}\left( {{{\widetilde \omega }_\alpha }{\rm{ - }}{{\widetilde \omega }_\beta } \pm \sqrt {{{\left( {{{\widetilde \omega }_\alpha }{\rm{ - }}{{\widetilde \omega }_\beta }} \right)}^2} + 4{{\rm{C}}_{\alpha \beta }}{{\rm{C}}_{\beta \alpha }}} ,2C_{\beta \alpha}} \right).
\end{equation}
The Q-factors of the two supermodes can be directly calculated from Eq.~(\ref{eigenfrequency}) through Eq.~(\ref{Q-factor}).

For a conservative system, it is required that $C_{\beta \alpha}=C_{\alpha \beta }^*$ and thus $C_{\beta \alpha}C_{\alpha \beta }$ is real~\cite{FRIEDRICH_Phys.Rev.A_interfering_1985-1,HEISS_Phys.Rev.E_repulsion_2000-1,WIERSIG_Phys.Rev.Lett._formation_2006-1,SONG_2010_Phys.Rev.Lett._Improving}. Then the coupling can be categorized as strong if $2|C_{\beta \alpha}|> |\widetilde\omega_{\alpha 2}-\widetilde\omega_{\beta 2}|$, or weak if $2|C_{\beta \alpha}|< |\widetilde\omega_{\alpha 2}-\widetilde\omega_{\beta 2}|$, with or without energy (level) repulsion (anti-crossing), respectively. When the system is open and not conservative with external coupling, generally $C_{\beta \alpha}C_{\alpha \beta }$ is not real and then it is not as direct as the conservative case to categorize the coupling anymore. But depending on whether there is effective anti-crossing of the energy, still the coupling can be roughly classified as strong or weak accordingly~\cite{WIERSIG_Phys.Rev.Lett._formation_2006-1,SONG_2010_Phys.Rev.Lett._Improving}.

For a passive and open system with two coupled quasi-bound modes, generally two new supermodes with different Q-factors arise.  In some parameter regimes the imaginary part of the eigenfrequency for one supermode can be greatly suppressed, leading to significant Q-factor enhancement that is possible in both strong and weak coupling regimes~\cite{FRIEDRICH_Phys.Rev.A_interfering_1985-1,HEISS_Phys.Rev.E_repulsion_2000-1,WIERSIG_Phys.Rev.Lett._formation_2006-1,SONG_2010_Phys.Rev.Lett._Improving}. This effect has been observed in many previous studies (see references in the main letter), but why the connection between Q-factor enhancement and multipolar conversions from lower to higher orders we reveal has been largely overlooked?

\begin{figure}[tp]

\centerline{\includegraphics[width=4cm]{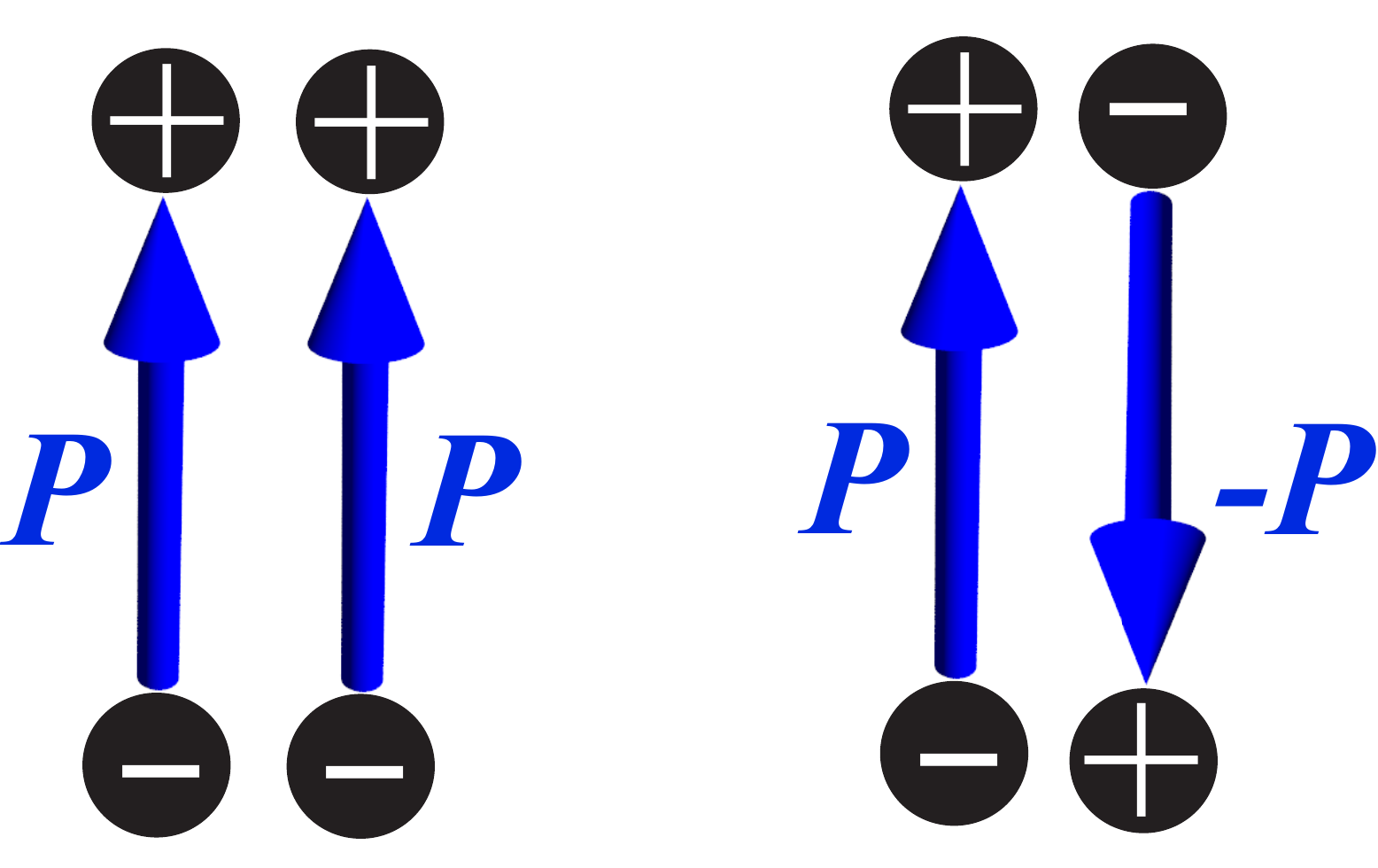}} \caption {\small Two ways to hybridize two parallel vertically orientated electric dipoles: in-phase hybridization induced supermode with lower Q-factor (left) and out-of-phase hybridization induced supermode with higher Q-factor (right). }\label{figr1}
\end{figure} 
\begin{figure*}[tp]
\centerline{\includegraphics[width=16cm]{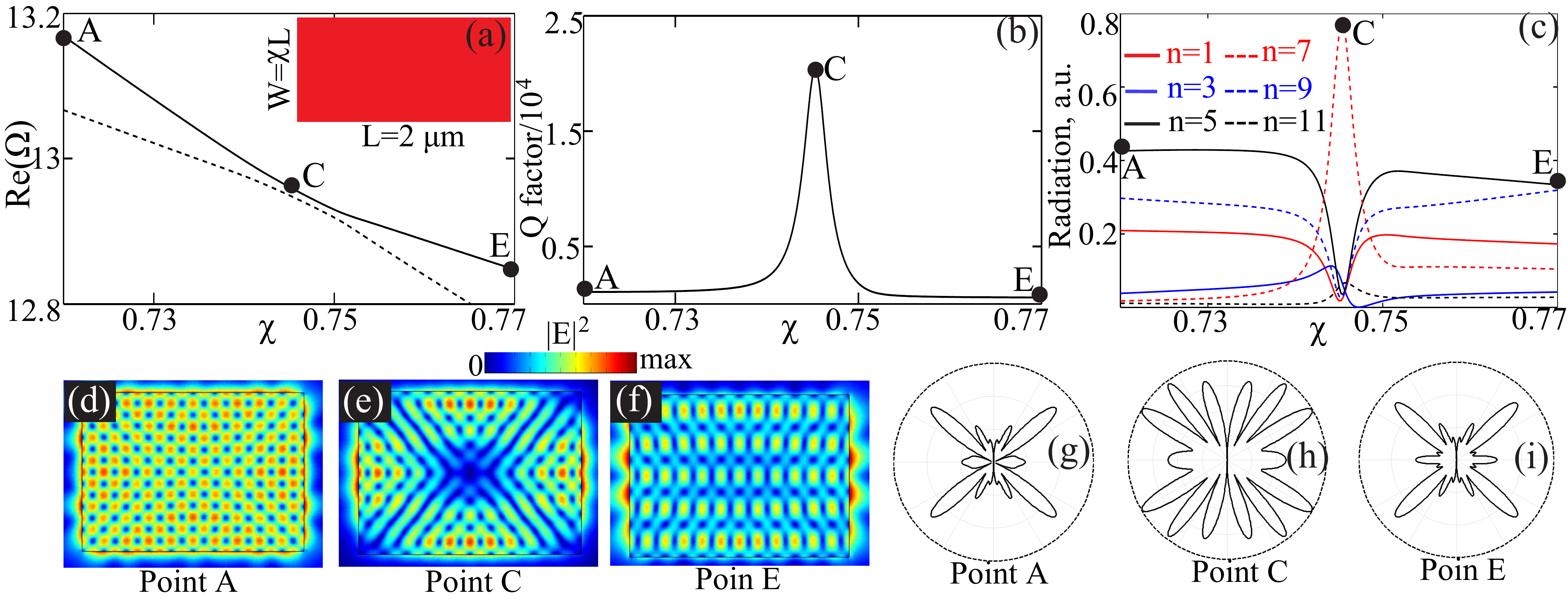}} \caption {\small The resonator studied is schematically shown in the inset of (a): a rectangular dielectric cavity with refractive index $3.3$, length $L=2~\mu$m and width to length aspect ratio $\chi=W/L$. (a) The dependence of real parts of the resonant frequencies for TE modes on the aspect ratio $\chi$. The coupling induces two supermode branches, and the solid (upper) and dashed (lower) curves correspond to the branches that experience significant Q-factor enhancement and suppression in the anti-crossing region, respectively.   Three supermodes are indicated by points $\textbf{A}$($\chi_{\rm{\textbf{A}}}=0.72$) , $\textbf{C}$ ($\chi_{\rm{\textbf{C}}}=0.7453$), and $\textbf{E}$ ($\chi_{\rm{\textbf{E}}}=0.77$), and the corresponding near-field (electric field intensity) and far-field (radiation pattern) distributions  of those supermodes are shown  accordingly in (d)-(i). (b) Evolutions of Q factors, and (c) the radiated power from all multipoles of order $n$ that are not negligible, for the supermodes located on the upper branch in (a). Results shown in sub-figures of (a), (d)-(f) are recalculated from Ref.~\cite{WIERSIG_Phys.Rev.Lett._formation_2006-1}.}\label{figr2}
\end{figure*} 

The core problem here is that the conventional two mode coupling model is kind of a scalar model. By saying so, we do not mean that in this model we treat the coupled modes as scalar waves, but rather that during the whole process of implementing the model shown above, we do not care much about the details of the vector field distribution of each mode. Indeed we would consider the vector nature of the coupled electromagnetic modes for the calculation of the off-diagonal coupling coefficients. Nevertheless, after writing down the interaction matrix shown in Eq.~(\ref{matrix}), the Q factors of the supermodes can be directly calculated without minding whether the involved modes are of vector, scalar or spinor nature (generally two wave functions $\mathbf{\Psi}_{\alpha,\beta}$ would be simply assigned). This is understandable as the Q factor is an overall feature of a mode, and it reflects the total energy dissipation rate but not directly linked to how the dissipation is distributed among different angular directions.

Based on this more or less scalar treatment, on one hand, the Q-factor enhancement can be rather simply attributed to the radiation loss suppression as a result of far-field destructive interference between $\mathbf{\Psi}_{\alpha,\beta}$ [the two wave functions would be superimposed according to the eigenvectors shown in Eq.~(\ref{eigenvector}), be the coupling strong or weak]. While on the other hand, however, the description of \textit{destructive interference} between two modes is itself, to some extent still hazy, but rather we do not know exactly what really happens in the far field that has induced the significant Q-factor enhancement. So the basic information here is that the connection we have revealed lies on the multipoles of vector nature, which is excluded from our kind of scalar treatment of the two mode coupling model.

$\blacksquare$~~~\textsc{\textbf{Perspective of vector mode hybridization and limitations}}\\
Now that the connection we have revealed is beyond the grasp of the conventional more or less scalar treatment, how about that we now pay full attention to the vector field nature and then investigate how their superposition would provide insight into the multipolar conversion induced Q-factor enhancement? Certainly this perspective can somehow explain the evolutions in the far field,  since far-field radiations and near-field charge-current distributions are not segregated but rather connected with each other through Maxwell equations. The coupling between the two modes would reshape the charge-current distributions in the near field and thus consequently also alter the far-field radiation patterns. But still, though valid, this is too general and to some extent obscure. The core question is:  will it be able to provide an intuitive and concrete physical picture to clarify why the Q-factor enhancement is subtly connected with multipolar conversions from lower to higher orders? The answer seems to be YES at the first glance: if we look at the simplest case of two coupled electric dipoles schematically shown in Fig.~\ref{figr1}, it is clear the out-of-phase hybridization induces effectively a higher-order electric quadrupole with the Q-factor enhanced compared to that of the uncoupled electric dipole~\cite{CHERN_2007_Phys.Rev.E_Particle}, which agrees exactly with what we have revealed; similar rules can be easily extended to coupling between higher order multipoles or multipoles of different orders.

Nevertheless, a thorough reexamination immediately spells some extra problems: the connection we have revealed is relying on multipoles that are represented by harmonics (spherical harmonics for 3D cases and cylinder harmonics for 2D cases), which are also termed as radiation multipoles~\cite{PAPASIMAKIS_NatMater_electromagnetic_2016}. In the quasi-static limit (charge-current distributed within a region far smaller than the effective wavelength), those radiation multipoles can be reduced to the electric and magnetic multipoles about which we have clear physical pictures: electric dipole corresponds to a pair of displaced charges of opposite signs (see Fig.~\ref{figr1}); magnetic dipole corresponds to a closed current loop; and so on and so forth. Under those circumstances, the vector hybridization perspective indeed can somehow explain intuitively the connection we have revealed. However, in the regimes beyond the quasi-static approximation (such as high index systems or systems of large geometric sizes), the radiation multipoles cannot be represented by simple charge-current distributions. That is to say, we name the spherical or cylindrical harmonic of order n as a multipole of the related order due to limited correspondence in the quasi-static regime. In general what we call a radiating multipole in the far field does not have to correspond to some real near-field ``multi-pole" charge-current distributions.  A compelling example of this is that a radiating electric dipole can correspond to a conventional electric dipole we know well (see Fig.~\ref{figr1}), or the less well known toroidal dipole~\cite{PAPASIMAKIS_NatMater_electromagnetic_2016}, or more complicated charge-current distributions (known as higher order corrections) that we do not have any intuitive physical picture about.  It is already very difficult to imagine what the hybridization of two toroidal dipole would lead to, let alone other higher order counterparts.

As a result, the vector hybridization model is helpful to clarify our finding for very limited cases in the quasi-static regime,  while fails to provide an intuitive and comprehensive basis for our more general studies. Basically the two mode coupling model can provide a hint about our discovered connection, but the subtleties of it are still exclusive.

$\blacksquare$~~~\textsc{\textbf{Other systems with multipolar conversion induced Q-factor enhancement in both strong and weak coupling regimes}}\\
After clarifying why the connection we have revealed is largely neglected in previous studies that employ the two mode coupling model in a more or less scalar way and how the vector hybridization model could only provide partial intuitive justification for such a connection, now we proceed to provide further evidence with sharply different two-mode coupling systems (in both strong and weak coupling regimes) to confirm that our discovery is widely applicable rather than being parasitic to specific geometries.

\begin{figure}[tp]
\centerline{\includegraphics[width=8.5cm]{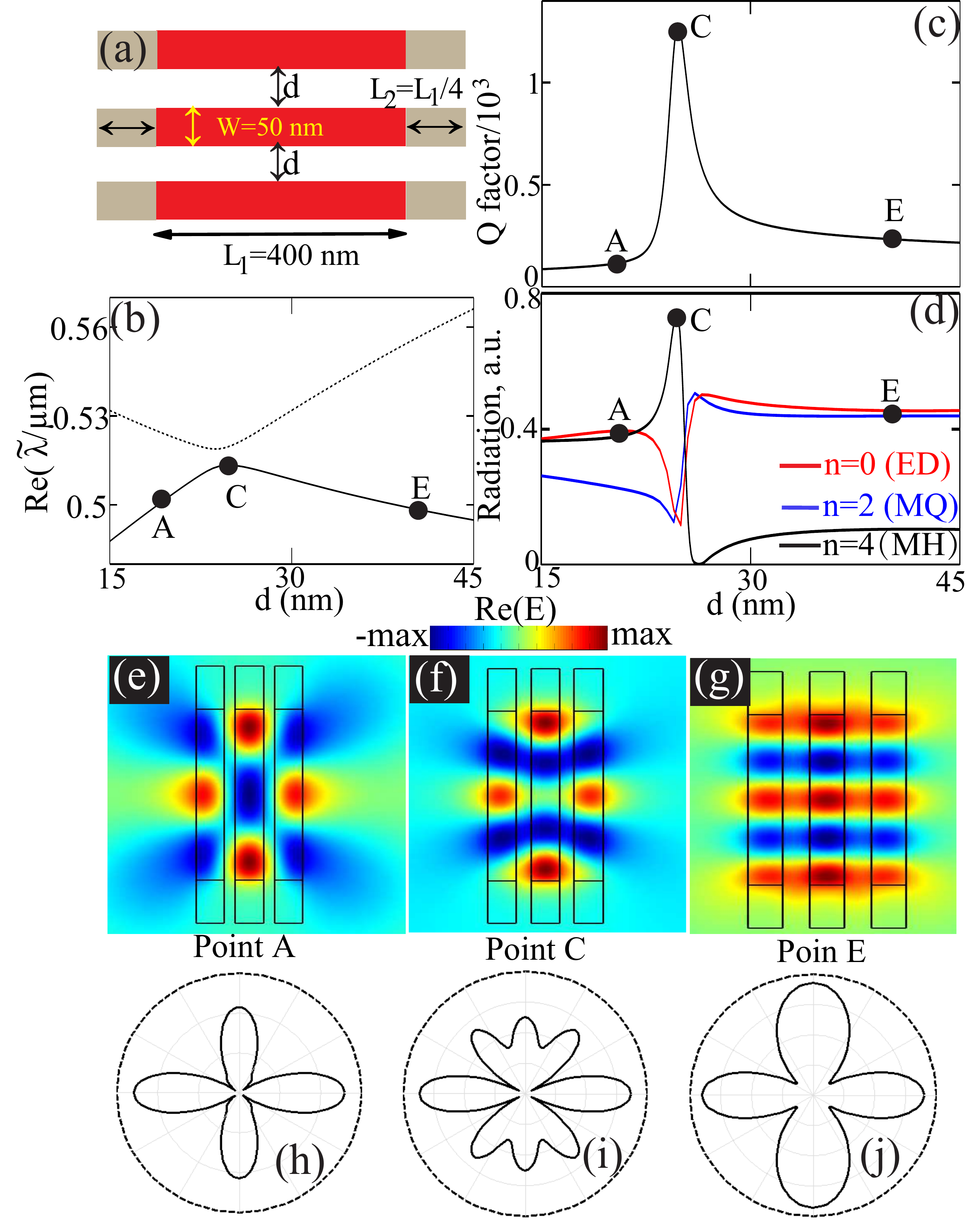}} \caption {\small (a) Schematic of a coupled system with threee identical dielectric (refractive index $3.5$) bars (length $L_1=400$~nm and width $W=50$~nm) coated by identical silver layers (length $L_2=100$~nm). The gaps between the bars are of the same width $d$. (b) The dependence of real parts of the resonant wavelengths for TM modes on the gap distance $d$. The coupling induces two supermode branches, and the solid (lower) and dashed (upper) curves correspond to the branches that experience significant Q-factor enhancement and suppression in the anti-crossing region, respectively. Three supermodes are indicated by points $\textbf{A}$ ($d_{\rm{\textbf{A}}}=20$~nm) , $\textbf{C}$ ($d_{\rm{\textbf{C}}}=25$~nm), and $\textbf{E}$ ($d_{\rm{\textbf{E}}}=40$~nm), and the corresponding near-field (out of plane electric field) and far-field (radiation pattern) distributions  of those supermodes are shown  in accordingly (e)-(j). (c) Evolutions of Q factors, and (d) the radiated power from all multipoles of order $n$ for the supermodes located on the lower branch in (b). Results shown in sub-figures of (b)-(c) and (e)-(g) are recalculated from Ref.~\cite{SONG_2010_Phys.Rev.Lett._Improving}.}\label{figr3}
\end{figure} 
\begin{figure}[tp]
\centerline{\includegraphics[width=8.5cm]{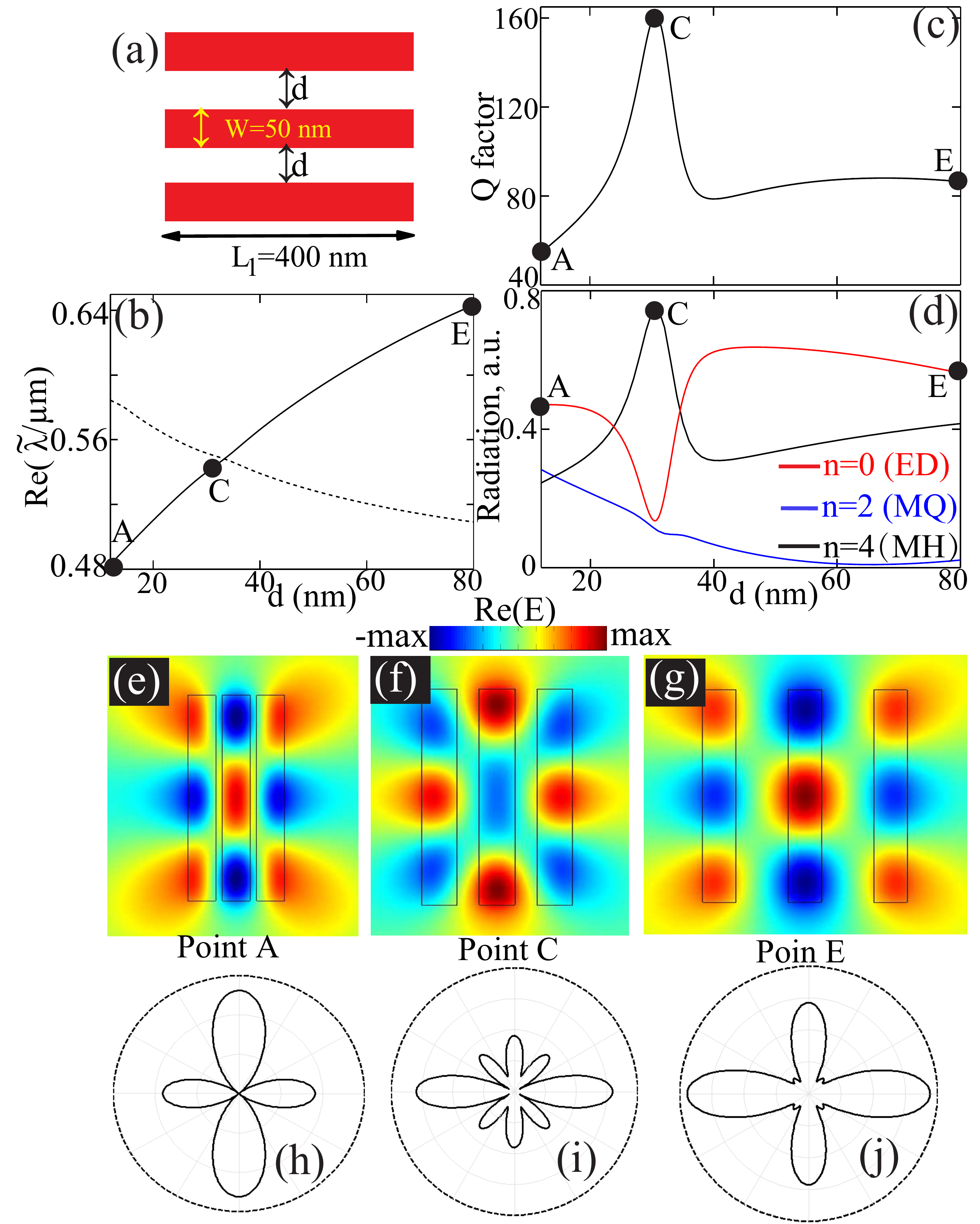}} \caption {\small (a) Schematic of a coupled system that is the same as that shown in Fig.~\ref{figr3}(a), excpet for that here all coating metal layers are removed. (b) The dependence of real parts of the resonant wavelengths for TM modes on the gap distance $d$. The coupling induces two supermode branches, and the solid  and dashed  curves correspond to the branches that experience significant Q-factor enhancement and suppression in the crossing region (weak coupling), respectively. Three supermodes are indicated by points $\textbf{A}$ ($d_{\rm{\textbf{A}}}=12$~nm) , $\textbf{C}$ ($d_{\rm{\textbf{C}}}=31$~nm), and $\textbf{E}$ ($d_{\rm{\textbf{E}}}=80$~nm), and the corresponding near-field (out of plane electric field) and far-field (radiation pattern) distributions are shown accordingly in (e)-(j). (c) Evolutions of Q factors, and (d) the radiated power from all multipoles of order $n$  for the supermodes located on the solid curve branch in (b). Results shown in sub-figures of (b)-(c), (e)-(g) are recalculated from Ref.~\cite{SONG_2010_Phys.Rev.Lett._Improving}}\label{figr4}
\end{figure} 

The first system we study is a homogeneous rectangular cavity (with refractive index $3.3$, length $L=2~\mu$m and width $W=\chi L$) schematically shown in the inset of Fig.~\ref{figr2}(a), a scenario that has already been studied in detail in Ref.~\cite{WIERSIG_Phys.Rev.Lett._formation_2006-1}. The spectra of the normalized resonant frequency $\Omega=\widetilde \omega L/c$ (real part) for TE modes (with on plane electric fields) with respect to aspect ratio $\chi$ are shown in Fig.~\ref{figr2}(a), where $\widetilde \omega$  and $c$ is the resonant frequency and speed of light, respectively. Distinct features of two mode coupling and energy (level) repulsion (anti-crossing) are exhibited, with three points indicated by \textbf{A}, \textbf{C} and \textbf{E} on the upper supermode branch.  Next we investigate in detail the process of Q-factor enhancement for the upper supermode branch, with the results shown in Fig.~\ref{figr2}(b). The Q factor is significantly enhanced in the anti-crossing region and reaches the optimum point (indicated by \textbf{C}) with maximum Q-factor. The corresponding results (radiated power from each contributing multipole with the total radiated power normalized) obtained through multipolar expansions of supermodes are shown in Fig.~\ref{figr2}(c), where the radiated power of the multipoles that are not shown is negligible. It is clear that the Q-factor enhancement is accompanied by the multipolar conversions from lower to higher orders (at point \textbf{C} the $n=7$ term is  dominant; while at points \textbf{A} and \textbf{E}, there are considerable lower order terms of $n=1$ and $n=5$; only odd-order terms are present due to the asymmetric nature of this supermode branch). To further verify the results from multipolar expansions, for the three supermodes indicated, the near-field intensity distributions and far-field radiation patterns are shown correspondingly in Figs.~\ref{figr2}(d)-(i).  At $\textbf{C}$ point there are more major radiation lobes than at the other two points, confirming the conversions between multipoles of different orders shown in Fig.~\ref{figr2}(c).

 Though exactly the same resonator has been studied in Ref.~\cite{WIERSIG_Phys.Rev.Lett._formation_2006-1}, two points that agree with our aforementioned statements should be emphasized: (i) In Ref.~\cite{WIERSIG_Phys.Rev.Lett._formation_2006-1} the two mode coupling model is employed in a sort of scalar way (pay no attention to the vector field distributions in the far-field) and thus the connection between Q-factor and multipolar conversion has not been identified (see sub-section of \textsc{Two mode coupling model and the Hidden connection}); (ii) Moreover, though the near-field hybridization views [shown in Figs.~R\ref{figr2}(d)-(f)] can explain the Q-factor enhancement due the destructive interference around the corners, they fail to provide a direct link to the multipolar conversions we have identified. This is understandable, as the near field distributions cannot tell intuitively or directly what the corresponding far-field multipoles are (see sub-section of \textsc{Perspective of vector mode hybridization and limitations}). This is more or less also the case for the two other scenarios that will be investigated below.

The next system we study is schematically shown in Fig.~\ref{figr3}(a), a case that has been studied in Ref.~\cite{SONG_2010_Phys.Rev.Lett._Improving} (refer to the caption of Fig.~R\ref{figr3} for parameter details and the silver is characterized by the Drude model: $\varepsilon (\omega ) = 3.92 - {{\omega _p^2} \mathord{\left/
 {\vphantom {{\omega _p^2} {{\omega ^2}\left( {1 + i\gamma } \right)}}} \right.\kern-\nulldelimiterspace} {{\omega ^2}\left( {1 + i\gamma } \right)}}$, with plasmon frequency $\omega _p=1.33\times10^{16}$ Hz and damping rate $\gamma=0.002$). The spectra of the resonant wavelength $\widetilde \lambda=\widetilde \omega c/2\pi$ (real part) for TM modes (with on plane magnetic fields) with respect to gap distance $d$ are show in Fig.~\ref{figr3}(b). Similar to what is shown in Fig.~\ref{figr2}, distinct features of two mode coupling and energy (level) repulsion are exhibited, with three points indicated by \textbf{A}, \textbf{C} and \textbf{E} on the lower supermode branch.  The process of Q-factor enhancement is shown Fig.~\ref{figr3}(c) and it is significantly enhanced in the anti-crossing region and reaches the optimum point (indicated by \textbf{C} point) with maximum Q-factor. The corresponding results (radiated power from each contributing multipole with the total radiated power normalized) obtained through multipolar expansions of supermodes are shown in Fig.~\ref{figr3}(d). In a similar fashion, Q-factor enhancement is intrinsically accompanied by the multipolar conversions from lower to higher orders: at point \textbf{C} the $n=4$ term (magnetic hexadecapole, MH) is  dominant; while at points \textbf{A} and \textbf{E}, there are considerable lower order terms of $n=0$ (electric dipole, ED) and $n=2$ (magnetic quadrupole, MQ); only even-order terms are present due to the symmetric nature of this supermode branch. For the three supermodes indicated, the near-field intensity distributions and far-field radiation patterns are shown correspondingly in Figs.~\ref{figr3}(e)-(j).

Now it is clear that the connection we have discovered pervades all the three sharply different scenarios studied (including those shown in Figs.~\ref{figr2} and \ref{figr3} and the one in the main letter). The common feature of all those cases is that there is very distinct energy (level) repulsion (anti-crossing of the real parts of eigenfrequencies), which originates from the strong mode coupling~\cite{HEISS_Phys.Rev.E_repulsion_2000-1,WIERSIG_Phys.Rev.Lett._formation_2006-1}. Then it is natural to ask another question: Will the connection we reveal be present in other weak coupling cases, that is, there is no energy repulsion. The weak coupling can be realized with a coupled system shown in Fig.~\ref{figr4}(a), which is the same as that shown in Fig.~\ref{figr3}(a), except that here the silver coatings are removed (also has been studied in Ref.~\cite{SONG_2010_Phys.Rev.Lett._Improving}). The spectra of the resonant wavelength shown in Fig.~\ref{figr4}(b) clearly indicate that there is no energy (level) repulsion, confirming its weak coupling nature, despite which there is still significant Q-factor enhancement for one supermode branch [Fig.~\ref{figr4}(c)]. The results of multipolar expansion for this supermode branch are shown in Fig.~\ref{figr4}(d), which confirms that the Q-factor enhancement in the weak coupling case is also accompanied by multipolar conversions from lower to higher orders: at point \textbf{C} the $n=4$ term (magnetic hexadecapole, MH) is  dominant; while at points \textbf{A} and \textbf{E}, there are considerable lower order terms of $n=0$ (electric dipole, ED) and/or $n=2$ (magnetic quadrupole, MQ); only even-order terms are present due to the symmetric nature of this supermode branch. This agrees well with the far-field radiation patterns shown in Figs.~\ref{figr4}(h)-(j)]. That is to say, the subtle connection we have discovered between significant Q-factor enhancement and multipolar conversions manifests itself in both strong and weak coupling regimes.

To summarize this section, we have provided concrete evidence confirming that the connection we reveal is quite general. We conclude the significance of our work as follows: (i) We have discovered a subtle connection that had been generally neglected in previous studies; (ii) Such a connection is quite general rather than being parasitic to specific systems, which pervades both strong and weak coupling regimes; (iii) The mechanism we have revealed is related to the fundamental feature (Q-factor of supermode) of a fundamental (two mode coupling) model in physics and thus our finding is of broad interest for many other involving fields, such as optomechanics, exciton-polaritonics, phonon-polaritonics and so on; (iv) As we have argued above, the current available model fails to provide a mathematically rigourous justification for our connection, thus naturally expecting that our work will stimulate lots of further work to improve the current model or to develop more advanced models to encapsulate all the subtleties of the general (two-modes or many-mode) coupling systems, benefiting all related fields.

\end{document}